\documentclass[11pt,a4paper]{amsart}

\usepackage[margin=1.5cm]{geometry}
\usepackage{amsmath,amsfonts,amssymb,amscd,amsthm}
\usepackage[usenames,dvipsnames,svgnames]{xcolor}

\usepackage{ucs}
\usepackage[T1]{fontenc}
\usepackage[utf8x]{inputenc}

\usepackage{graphicx}
\graphicspath{{figures/}}
\usepackage{caption}
\usepackage[labelformat=simple]{subcaption}

\usepackage[numbers]{natbib}
\usepackage{doi}
\usepackage{autonum}
\usepackage{enumitem}
\usepackage{makecell}
\usepackage{hyperref}
\usepackage{acronym}
\usepackage{listings}
\usepackage{textgreek}
\usepackage[newfloat,frozencache]{minted}
\usepackage{amsmath}
\usepackage{hyperref}
\usepackage{url}
\usepackage[colorinlistoftodos, textwidth=1.5\marginparwidth, shadow]{todonotes}
\usepackage{color}
\usepackage{framed}
\usepackage{verbatim}
\usepackage{fancyvrb}
\usepackage{multicol}
\usepackage{tcolorbox}
\usepackage{booktabs}

\definecolor{shadecolor}{gray}{.92}
\definecolor{incolor}{rgb}{0,0,.7}
\definecolor{outcolor}{rgb}{.65,0,0}
\definecolor{syntaxcolor}{rgb}{.65,0,0}
\definecolor{bg}{rgb}{0.93,0.93,0.93}
\definecolor{myblue}{RGB}{93,188,210}
\definecolor{mygreen}{RGB}{189,210,93}
\definecolor{myorange}{RGB}{210,173,93}
\definecolor{myred}{RGB}{210,93,130}
\definecolor{mydarkblue}{RGB}{93,130,210}
\definecolor{mydarkgreen}{RGB}{93,210,173}

\acrodef{dof}[DOF]{Degree Of Freedom}
\acrodefplural{dof}[DOFs]{Degrees Of Freedom}
\acrodef{dg}[DG]{Discontinuous Galerkin}
\acrodef{rt}[RT]{Raviart-Thomas}
\acrodef{fe}[FE]{Finite Element}
\acrodefplural{fe}[FEs]{Finite Elements}
\acrodef{jit}[JIT]{just-in-time}
\acrodef{ode}[ODE]{Ordinary Differential Equation}
\acrodef{pde}[PDE]{Partial Differential Equation}
\acrodefplural{pde}[PDEs]{Partial Differential Equations}
\acrodef{api}[API]{Application Programming Interface}
\acrodef{dsl}[DSL]{Domain-Specific Language}
\acrodefplural{dsl}[DSLs]{Domain-Specific Languages}
\acrodef{amg}[AMG]{Algebraic MultiGrid}
\acrodef{hpc}[HPC]{High Performance Computing}

\theoremstyle{definition}

\newcommand{\thetitle}{The software design of Gridap: a Finite Element package\\ based on the Julia JIT compiler}
\newcommand{\theshortitle}{The software design of Gridap}

\newcommand{\sbc}[1]{{\color{blue}{#1}}}

\newcommand{\fig}[1]{Fig.~\ref{#1}}

\newcommand{\sect}[1]{Sect.~\ref{#1}}

\newcommand{\lst}[1]{Listing~\ref{#1}}

\newcommand{\equ}[1]{Eq.~\eqref{#1}}
\newcommand{\tabl}[1]{Table~\ref{#1}}

\newcommand{\juliacode}[1]{
\includegraphics[width=\textwidth]{#1.pdf}
\vspace*{-1.5em}
}

\newcommand{\jl}[1]{\small\Verb{#1}}
\newcommand{\jll}[1]{\texttt{#1}}

\setlist[description]{leftmargin=1em,labelindent=1em,noitemsep,topsep=0pt}

\hypersetup{
    bookmarks=true,
    breaklinks=true,
    bookmarksopen=true,
    pdftitle={\thetitle},
    pdfauthor={Francesc Verdugo and Santiago Badia},
    colorlinks=true,
    linkcolor=black,
    citecolor=blue,
    filecolor=black,
    urlcolor=blue
}

\newcounter{julialoc}
\tikzstyle{nodest}=[shape=rectangle, minimum width=1.5em, minimum height=1.5em, rounded corners, align=center, fill=white]
\tikzstyle{leafst}=[shape=rectangle, minimum width=1.5em, minimum height=1.5em, rounded corners, align=center, fill=mygreen!90]

\tikzstyle{edge from parent}=[draw,line width=1.5pt,black!50]

\begin{document}

\title[\theshortitle]{\thetitle}

\author[F. Verdugo]{Francesc Verdugo$^{1,*}$}
\author[S. Badia]{Santiago Badia$^{1,2}$}

\thanks{\null\\
$^1$ CIMNE, Centre Internacional de M\`etodes Num\`erics a l'Enginyeria, Esteve Terrades 5, E-08860 Castelldefels, Spain.\\
$^2$ School of Mathematics, Monash University, Clayton, Victoria, 3800, Australia.\\
$^*$ Corresponding author.\\
E-mails: {\tt fverdugo@cimne.upc.edu} (FV), {\tt santiago.badia@monash.edu} (SB)}

\date{\today}

\begin{abstract}
We present the software design of Gridap, a novel finite element library written exclusively in the Julia programming language, which  is being used by several research groups world-wide to simulate complex physical phenomena such as magnetohydrodynamics, photonics, weather modeling, non-linear solid mechanics, and fluid-structure interaction problems. The library provides a feature-rich set of discretization techniques for the numerical approximation of a wide range of mathematical models governed by Partial Differential Equations (PDEs), including linear, nonlinear, single-field, and multi-field equations. An expressive API allows users to define PDEs in weak form by a syntax close to the mathematical notation. While this is also available in previous frameworks, the main novelty of Gridap is that it implements this API without introducing a domain-specific language plus a compiler of variational forms. Instead, it leverages the Julia just-in-time compiler to build efficient code, specialized for the concrete problem at hand. As a result, there is no need to use different languages for the computational back-end and the user front-end anymore, thus eliminating the so-called two-language problem. Gridap also provides a low-level API that is modular and extensible via the multiple-dispatch paradigm of Julia and provides easy access to the main building blocks of the library if required. The main contribution of this paper is the detailed presentation of the novel software abstractions behind the Gridap design that leverages the new software possibilities provided by the Julia language. The second main contribution of the article is a performance comparison against FEniCS. We measure CPU times needed to assemble discrete systems of linear equations for different problem types and show that the performance of Gridap is comparable to FEniCS, demonstrating that the new software design does not compromise performance. Gridap is freely available at Github (github.com/gridap/Gridap.jl) and distributed under an MIT license.
\end{abstract}

\maketitle






\section*{Program summary}

\begin{itemize}
\item Program title: Gridap.jl (version 0.16)
\item Developer's respository link:  \url{https://github.com/gridap/Gridap.jl}
\item Licensing provisions: MIT license (MIT)
\item Programming language: Julia
\item Supplementary material: Source code of the Listings presented in this paper. Each Listing below indicates the name of its corresponding source file.
\item Nature of problem: Computational simulation of a broad range of application problems governed by partial differential equations including linear, nonlinear, single field, and multi-physics problems. Gridap is currently being used by several research groups world-wide to simulate complex physical phenomena such as magnetohydrodynamics, photonics, weather modeling, non-linear solid mechanics, and fluid-structure interaction problems.
\item  Solution method:  Arbitrary-order grad-, curl-, and div-conforming finite elements on n-cube and n-simplex meshes. Continuous and Discontinuous Galerkin methods. Newton–Raphson linearization. Krylov subspace iterative solvers. Sparse direct solvers.
\end{itemize}

\section{Introduction}

Mathematical models based on \acp{pde}
 have gained popularity thanks to the computational power increase unlocked by modern computers, which has pushed forward the complexity of the problems engineers and scientists can simulate. 
However, computational power alone is not enough. Recent advances in numerical discretizations, e.g., \ac{fe} methods, and efficient implementation are required to efficiently and accurately solve \acp{pde} in the most complex scenarios. Besides, domain-specific researchers and practitioners that need to simulate challenging applications in their fields are not necessarily programming and numerical experts. Thus, there is a need for software packages that provide high-performance state-of-the-art simulations tools with simple user interfaces. As a consequence, a myriad of \ac{fe} frameworks have been proposed, from libraries designed to solve a narrow set of problems with specific computational techniques to general-purpose \ac{fe} packages that provide a rich set of discretization methods and can solve a wide spectrum of \acp{pde}.

The development of numerical frameworks is a huge endeavor. Since numerical techniques share many software components, it is wise to leverage this effort by broadening the scope without incurring performance hits. However, the design and development of flexible and extensible general-purpose \ac{fe} packages is not an obvious task. Several software designs have been proposed with this goal in mind. On the one hand, \ac{fe} libraries like Deal.II \cite{Bangerth2007}, FEMPAR \cite{Badia2018}, and MoFEM \cite{Kaczmarczyk2020} provide the building blocks that are common in different types of \ac{fe} solvers,  allowing users to combine them for their particular needs. These libraries are usually modular and extensible via an object-oriented design that provides flexibility to accommodate new scenarios. However, working with these libraries involves a steep learning curve and can be tedious. These libraries are written in static languages like C/C++ or Fortran to provide performance. The user needs to write a considerable amount of glue code to put all pieces together (e.g., one typically must write the \ac{fe} assembly loop by hand). The \ac{api} is usually also in a compiled language to eliminate inter-language boundaries when combining the underlying tools, which further worsens the user experience.

An alternative approach that aims at solving these pitfalls are \ac{fe} frameworks like FEniCS/DOLFIN \cite{Logg2010}, Firedrake \cite{Rathgeber2016}, and FreeFEM \cite{free_fem_web}. These libraries are written in compiled languages to be efficient but allow the user to define the \ac{fe} problem with a high-level \ac{dsl} that is compact and mimics the underlying mathematical definition of the weak form of the \ac{pde}. The code takes the high-level definition of the problem and automatically generates an efficient \ac{fe} code, e.g., releasing the user from writing for loops in the \ac{fe} assembly. Moreover, this high-level \ac{dsl}  is usually available in a dynamic language, typically Python, facilitating the usage of the library and its coupling with other software components from, e.g., the Python ecosystem. Unfortunately, this approach has also its drawbacks. Performance is achieved by a compiler of variational forms, e.g., FFC \cite{Kirby2006}. It generates efficient low-level code in C/C++ that is specialized for the concrete problem defined via the high-level \ac{api}. This approach is affected by a two-language (or even three-language, counting the \ac{dsl}) barrier. The user workspace is limited, and its extension is not obvious. This introduces rigidity when it comes to extending the library since form compilers are sophisticated systems not usually designed to be extended by average users.

In this paper, we present the software design of Gridap \cite{Badia2020}, a novel, free, and open-source \ac{fe} library distributed under a MIT license that aims at solving the drawbacks of previous packages while keeping their strengths. That is, we aim at combining the extensibility and composability of libraries like Deal.II with the ease of use and compact high-level syntax of frameworks like FEniCS without compromising run-time performance. The novel Gridap software design leverages the capabilities of the emerging Julia programming language \cite{Bezanson2017}. Julia provides the performance of compiled computer languages like C/C++ or Fortran and the productivity of interpreted ones like Python or MATLAB. As a result, there is no need to use different languages for the computational back-end and the user front-end anymore, thus eliminating the so-called \emph{two-language problem}. This is accomplished by the Julia \ac{jit} compiler, which parses high-level code (similar to Python and MATLAB) and generates efficient native machine code that is specialized for the particular types encountered at run-time, thus being as efficient as statically compiled code implemented in C/C++ or Fortran. The main design goal of Gridap is to exploit the Julia \ac{jit} compiler to construct a modern \ac{fe} library that is easy to use and maintain and is also efficient. In particular, we use the \ac{jit} compiler to generate problem-specific native machine code as an alternative to developing and maintaining a compiler of variational forms like FFC at the package level. The result is a library with a compact high-level \ac{api} similar to FEniCS but much easier to maintain and extend. Gridap also provides a modular low-level \ac{api} that gives access to the main building blocks of the library if required, thus providing expressive high-level and a low-level interface at the same time. This low-level \ac{api} is also coded in Julia. It is accessible to users, which can easily extend the library via the multiple-dispatching paradigm in Julia. In contrast, the extension of deal.II, FEMPAR or FEniCS requires modifying the core of the library, written in C/C++ or Fortran. Gridap can also be easily coupled with other packages from the Julia ecosystem, e.g., for optimization \cite{Dunning2017}, ordinary differential equations \cite{Rackauckas2017},  machine learning \cite{Innes2018},  automatic differentiation \cite{Revels2016}, or scientific visualization \cite{Danisch2021}. Being completely written in Julia, Gridap solvers can be automatically differentiated using automatic differentiation tools to compute, e.g., sensitivities of solvers to parameters.

In any case, Gridap is not a simple Julia translation of an existing \ac{fe} code. There are other more standard \ac{fe} libraries written in Julia, see, e.g., FinEtools \cite{finetools_gh}, JuliaFEM \cite{Frondelius2017}, or Ferrite \citep{ferrite_gh}, whose interface is inspired by Deal.II. In contrast, Gridap relies on a novel representation of data structures associated with the cells in the computational mesh. The user can manipulate cell-wise data without the need to explicitly iterate over the cells as required in other \ac{fe} packages. It leads to a much more compact and expressive interface while keeping memory requirements similar to traditional codes. Quantities are not stored for all mesh cells simultaneously. Instead, Gridap relies on so-called \emph{lazy} arrays that compute their entries on demand. {Combining these ideas with the Julia \ac{jit} compilation, Gridap provides a high-level API that resembles the whiteboard mathematical statement of weak forms of \acp{pde}.}

The main contribution of this paper is the detailed presentation of the novel software abstractions behind the Gridap design (as of version 0.16). Due to length constraints, we focus on the low-level and high-level tools used to build elemental matrices and vectors, which is the part of the library that follows a more unconventional software approach and is the cornerstone of the high-level \ac{api} close to the mathematical notation. We present both single-field and multi-field problems. Note however that Gridap is a feature-rich \ac{fe} library that supports a large set of discretization techniques that go beyond the ones covered in this paper (including different types of continuous and \ac{dg} methods) and supports a wide range of \acp{pde} types, including, e.g., nonlinear and complex-variable equations. The second main contribution of the article is a performance comparison against FEniCS. We measure CPU times needed to assemble discrete systems of linear equations for different problem types and show that the performance of Gridap is comparable to FEniCS, demonstrating that the new software design of Gridap does not compromise performance.

The next sections of the paper are structured as follows. \sect{sec:backend} presents the low-level software components of the Gridap project. At the end of this section, we show how one can combine these low-level tools to compute elemental stiffness matrices for all cells of the \ac{fe} mesh without explicitly writing for-loops. Then, in \sect{sec:frontend}, we show how the high-level interface wraps low-level tools, which allows one to manipulate quantities in the \ac{fe} computation in an even more convenient way. \sect{sec:multifield} is about the extension to multi-field \acp{pde}.  \sect{sec:examples} contains numerical experiments that show the performance of the library. We draw some conclusions in \sect{sec:conclusions}.

\section{The computational back-end}
\label{sec:backend}

\subsection{Overview}

We present the architecture of the computational back-end of Gridap with a bottom-up approach, starting from the lowest-level functionality and building on top of it. The back-end is structured in terms of several abstractions represented by \emph{abstract types}, which are the Julia counterparts of classes with virtual methods in C++. These abstractions make the code modular and extensible since users and developers can implement new specializations of these abstract interfaces if needed. The main abstract types are listed in \tabl{tabl:interfaces}. Note that this part of the code relies on very few abstractions. The complexity of a \ac{fe} solver is achieved by combining these basic building blocks. For instance, the local shape functions in a reference cell are represented as vectors of fields, thus combining the interfaces of the abstract types \jl{AbstractArray} and \jl{Field}, instead of introducing a new abstraction for this particular case. Note that the abstract types in \tabl{tabl:interfaces} stem naturally from fundamental mathematical concepts and are not introduced due to implementation artifacts. A minor exception to this rule is the abstract type \jl{Map}, which is introduced mainly for performance reasons. \jl{Map} is the fundamental abstraction that allows us to reuse computations and avoid spurious dynamic allocations when iterating over the cells of the mesh. This type could only be replaced by \jl{Function} if performance was not a concern. In the following subsections, we describe how we use and extend the \jl{AbstractArray} interface and how we introduce important performance optimizations via the \jl{Map} abstract type. We also detail the other abstract types listed in \tabl{tabl:interfaces}. Finally, we show how to build core components of a \ac{fe} solver combining these basic ingredients, e.g., to implement the local interpolation spaces or elemental matrices associated with a \ac{pde} weak form and the cells of a \ac{fe} mesh.

\begin{table}[ht!]
\begin{tabular}{llp{0.5\textwidth}}
\toprule
Defined in & Abstract type & Purpose\\
\midrule
Julia & \jl{Number} & Umbrella for all types representing numbers.\\
Julia & \jl{Function} &  Defines the \ac{api} of callable objects.  \\
Julia & \jl{AbstractArray} & Defines the \ac{api} of array-like objects, such as indexing, iteration, size, axes, etc. \\
Gridap & \jl{Map} & Defines an \ac{api} for callable objects that are able to reuse data between different calls. \\
Gridap & \jl{MultiValue} & Defines the \ac{api} of arbitrary rank tensors used to represent tensor-valued physical quantities.\\
	Gridap & \jl{Field} & Defines the \ac{api} of physical fields, i.e., functions that evaluated at a point in a physical domain return a scalar, vector or tensor value represented by some object of \jl{Number} or \jl{Multivalue} type.\\
Gridap & \jl{Dof} & Defines the \ac{api} of linear functionals over physical fields. I.e., a \jl{Dof} object can be called on a \jl{Field} object returning a scalar value.\\
\bottomrule
\end{tabular}
\caption{Summary of the main abstract interfaces of the Gridap back-end.}
\label{tabl:interfaces}
\end{table}

\subsection{Julia arrays}

Julia provides an excellent framework to work with multi-dimensional arrays similar to NumPy in Python. The main difference is that the user can write arbitrary for-loops over these arrays without performance limitations. Explicit loops are usually slightly faster than using broadcast operations, which might be surprising to Python programmers. Julia provides the built-in array type \jl{Array{T,N}}, which is a \emph{parametric} type with type parameters \jl{T} and \jl{N}, being the type of the elements stored in the array and the dimension of the array, respectively. In particular, vectors are represented by the type \jl{Array{T,1}} and matrices by the type \jl{Array{T,2}}, or by the aliases \jl{Vector{T}} and \jl{Matrix{T}} respectively.  If the element type \jl{T} has values contained in a chunk of bits (i.e., an \jl{isbits} type in Julia notation), e.g., \jl{Int32} or \jl{Float64}, the internal memory layout of the built-in array type is a one-dimensional contiguous section of memory.  The length is equal to the number of bits needed to represent one instance of type \jl{T} times the number of elements in the array.  Julia sweeps the array entries in \emph{row-major} order to represent multi-dimensional arrays with a 1D section of memory, like Fortran. The default Julia array type is a reference to a heap-allocated section of memory similar to a dynamically allocated array in C/C++ or Fortran. In particular, this means that a Julia array of Julia arrays is a reference to a vector of references pointing to the different memory locations storing the local arrays, similar to an array of pointers in C/C++.

In addition to the built-in array type \jl{Array{T,N}}, Julia provides the abstract type \jl{AbstractArray{T,N}} that represents types that behave like an array but are not necessarily the built-in array type. It allows one to implement new array types optimized for particular scenarios to achieve better performance. These new concrete types of  \jl{AbstractArray{T,N}} \emph{specialize} the abstract type by implementing its interface. To illustrate this, \lst{lst:arrays} shows how to implement a new vector type for the particular case of vectors with all entries having the same value.  Formally, we define the element $v_i$ of the constant vector as $v_i \doteq \hat v$ for all $i=1,\ldots,|v|$, where $\hat v$ is the unique value stored in the array and $|v|$ denotes the array length. This specialized vector implementation is less memory consuming than the default one.

\begin{listing}[ht!]
\juliacode{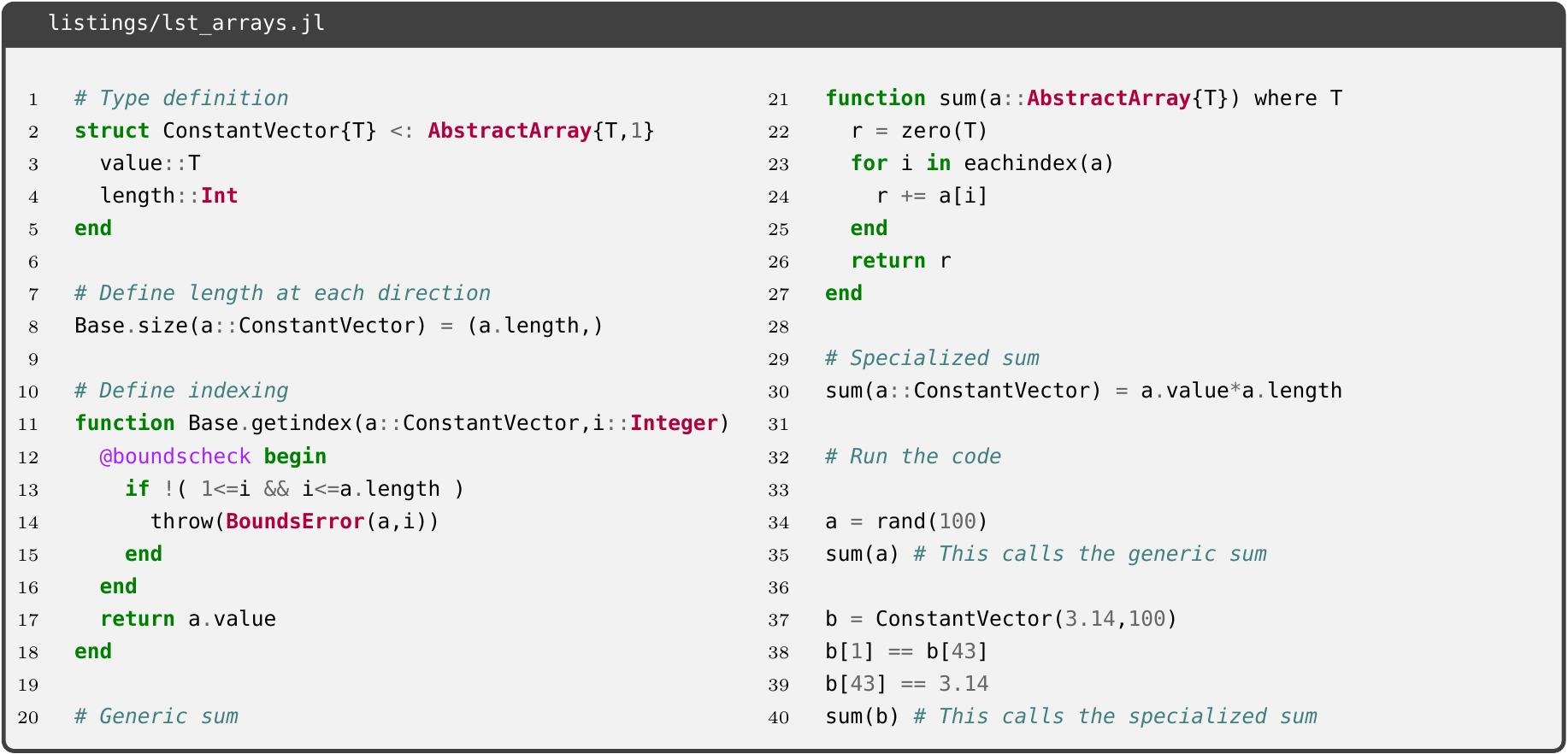}
\caption{Implementing a custom array type in Julia.}
\label{lst:arrays}
\end{listing}

One of the most powerful features of Julia is that one can write \emph{generic} code for any sub-type implementing the interface of an abstract type without suffering from the performance hit associated with polymorphic methods in object-oriented languages like C++, Fortran 2003, or Python. It is possible thanks to the Julia \ac{jit} compiler, which converts just before run-time the generic code into efficient machine code specialized for the concrete array types used in the computation. For instance, we implement a vanilla version of the built-in function \jl{sum} that sums all the entries in the given array in line 21 of \lst{lst:arrays}. When this function is called for a concrete array type, the Julia \ac{jit} compiler will compile a specialized version of this function for the concrete types it encounters in the argument list, resulting in efficient code. In addition, it is possible to help the \ac{jit} compiler to generate even more efficient machine code by providing specialized function definitions using so-called \emph{multiple dispatch}. For instance, the sum of vector entries does not require looping over all entries when they are all the same. Thus, the generic method is not as efficient as it could be even if it is specialized by the compiler. The developer can help the compiler by defining an optimized version that is dispatched  for this particular case as shown in line 30 of \lst{lst:arrays}. This strategy is called multi-dispatch since one can specialize any number of arguments simultaneously in the function definition.

\subsection{Main custom array types used in the Finite Element code}

Apart from the built-in Julia arrays, Gridap considers several other array types to leverage optimizations that apply in \ac{fe} computations. These arrays represent cell-wise data efficiently and conveniently. Different algorithms in the code are implemented in terms of these arrays. Custom array types allow one to manipulate cell-wise data independently of how it lays in memory. For instance, one can write a generic code in terms of an array representing the quadrature points for all mesh cells without exposing how these points are stored.  The same generic code can work in different scenarios, e.g., for meshes with a single reference element, where one can use a constant array similar to the one presented in \lst{lst:arrays} to store the quadrature points, and also for meshes with more than one element type that will require the usage of a more sophisticated array type, i.e., the so-called \jl{CompressedArray} introduced later. We avoid code duplication by using generic code and custom array types while keeping performance by using the Julia \ac{jit} compiler to specialize the emitted machine code for each case.

Gridap uses different array types in its implementation being \jl{Fill}, \jl{CompressedArray}, \jl{Table}, \jl{CachedArray}, and \jl{LazyArray} the most important ones. The type \jl{Fill} is a curated implementation of a constant multi-dimensional array provided in the Julia package FillArrays \cite{fillarrays_gh}. \jl{Fill} is similar to the constant vector presented in \lst{lst:arrays}, but it comes with some improvements like support for the multi-dimensional case.  \jl{CompressedArray} is a type implementing an array $v$, whose  entry $v_i$ is defined as $v_i\doteq \hat v_{j}$ with $j=t_i$. The quantity $\hat v_j$ represents a small vector (small with respect to $v$) containing the (unique) values of $v$ and $t_i$ is an indirection array that for each index $i$ returns the index $j$ in the small vector $\hat v_j$. E.g., this array type represents quantities defined in reference elements for \ac{fe} spaces with multiple element types. In this context, $\hat v_j$ can be interpreted as the reference value associated with element type $j$. $t_i$ can be interpreted as the element type for a cell id $i$. Note that the type \jl{CompressedArray} only stores in memory the vector of reference values $\hat v_j$ and the vector of cell types $t_j$. The next array type is \jl{Table}, which represents a vector of vectors in a memory-efficient way. E.g., this array stores cell connectivities (a vector of node ids for each cell id) in a general unstructured grid with potentially different element types (i.e., different number of local nodes per cell). This type adopts the usual representation of a \emph{jagged array}, i.e., vector of vectors, using a contiguous vector of values $a$ and an auxiliary vector of pointers $p$. The entry $v_i$, $i=1,\ldots,|v|$, of a \jl{Table} is a small vector consisting in a consecutive portion of the vector $a$ for indices delimited by the vector of pointers $p$, namely  $v_i=(a_r,a_{r+1}\ldots,a_s)^{\rm t}$ with $r\doteq p_i$ and $s\doteq p_{i+1}-1$. We use this data layout instead of a built-in Julia vector of Julia vectors since the data of this latter type is not stored consecutively in memory, thus not leveraging memory locality. In contrast, the data stored in the custom array type \jl{Table} is stored in memory consecutively, which contributes to more efficient memory access. Another important array type is \jl{CachedArray} which implements a multi-dimensional array that can change its size efficiently. E.g., this array type represents elemental matrices and vectors whose size can change between cells for \acp{fe} spaces with multiple element types. The size of an instance of \jl{CachedArray} changes by calling 
 \jl{setsize!(a,new_size)}; it sets the size of \jl{a}  to match the tuple of integers \jl{new_size}. The implementation of \jl{setsize!} allocates a new array under-the-hood each time a new size is requested. The allocated arrays are cached in this process so that it is not needed to allocate a new array if one with the right size was already cached. This caching strategy is designed having \ac{fe} computations in mind since the number of different array sizes one needs for an instance of \jl{CachedArray} is usually the same as the number of different element types in the mesh. Thus, the number of dynamic allocations when iterating over a mesh is proportional to the number of element types but independent of the total number of cells. The last array type, \jl{LazyArray}, deserves a more in-depth discussion and will be presented in \sect{sec:lazy_array}.

\subsection{Avoiding heap allocations when indexing arrays of arrays}

When implementing custom array types, like \jl{Table}, reducing the number of objects dynamically allocated in the heap is challenging. If the value returned when indexing an array \jl{a} at index \jl{i}, namely \jl{a[i]}, is an \jl{isbits} object (e.g., an instance of \jl{Int} or \jl{Float64}), the result will be efficiently allocated in the stack of the calling function and the programmer does not need to take any extra action related with memory allocation.  However, if \jl{a[i]} is not an \jl{isbits} object, e.g., if \jl{a[i]}  is an instance of a Julia \jl{Array{T,N}}, the result is usually allocated in the heap by the Julia compiler. In this case, dynamically allocating objects when fetching entries \jl{a[i]} while iterating over large arrays is not efficient. One needs to adopt a mechanism to cut down the number of dynamic allocations.  A naive solution is to pre-allocate the result of \jl{a[i]} and store it in a global variable. Then, when indexing \jl{a[i]} for a particular \jl{i}, one can fill the pre-allocated object and return it without allocating new memory dynamically. However, this naive solution is not the one adopted in Gridap because it is not thread-safe. Several threads cannot use the same pre-allocated output without incurring in race conditions.

To solve the problem of memory allocations in a thread-safe manner, Gridap introduces its own API to iterate over arrays, see \lst{lst:array_cache}. This API defines two new functions, namely \jl{array_cache} and \jl{getindex!}, to iterate arrays efficiently. Before starting the loop, one calls \jl{array_cache} to generate a cache object holding arbitrary data. This cache is reused each time the array is indexed. Function \jl{getindex!} is a generalization of the built-in function \jl{Base.getindex}, which accepts a cache object in its first argument that can speed up the computation of the result.  Custom array types overload  \jl{array_cache} and \jl{getindex!} to establish which data is reused and how it is reused at each iteration. For instance, the cache object can store a pre-allocated object that is filled and returned when calling \jl{getindex!}, thus avoiding allocating new objects within the loop. Note that this strategy is thread-safe; one can generate an independent cache for each thread before starting the loop (see, e.g. the last loop in \lst{lst:array_cache}), thus avoiding race conditions.

\begin{listing}[ht!]
\juliacode{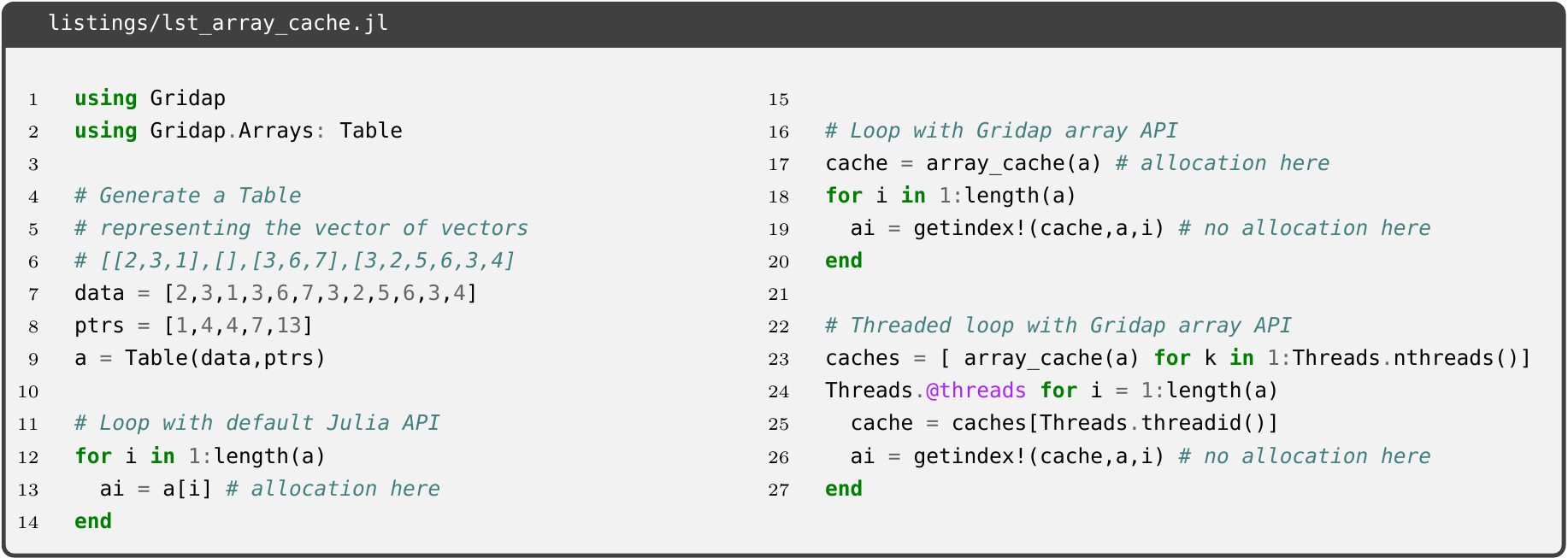}
\caption{Demonstration of the Gridap array API that allows to iterate over arrays without allocating memory in the heap at each iteration.}
\label{lst:array_cache}
\end{listing}

The new Gridap API for looping over arrays is general. It can be used to iterate over any array type specializing \jl{AbstractArray}, including (but not limited to) the array types defined in Gridap. By default, \jl{array_cache} does nothing and \jl{getindex!} calls the built-in function \jl{Base.getindex}. Thus, using the Gridap API to iterate arrays that do not define the functions \jl{array_cache} and \jl{getindex!} is equivalent to iterating them with the standard Julia API. It means that one does not need to define \jl{array_cache} and \jl{getindex!} for array types that do not need to pre-allocate any output or reuse any data.

\subsection{Mapping large arrays under memory constraints}
\label{sec:lazy_array}

{Many \ac{pde} discretization techniques, e.g., \ac{fe} methods, can be expressed as complex cell-wise operations between fields, linear functionals, scalars and tensors. Here, we provide expressive and efficient machinery for these cell-wise operations. We use the FE method as a driving application. However, the software abstractions are general in scope and apply to other scenarios with similar requirements.}

Like other programming languages, Julia provides the function \jl{map}, which transforms arrays and other collections in a very convenient way via a high-level API without the need of explicitly writing for-loops.  When applied to arrays, the \jl{map} function returns an array $r$, whose entries are computed by applying a given function $f$ entry-by-entry to a number $n$ of input arrays $a^1,\ldots,a^n$, namely $r_i \doteq f(a^1_i,\ldots,a^n_i)$. The syntax provided by the \jl{map} function is especially convenient to implement \ac{fe} codes since it allows one to represent quantities defined on the cells of a \ac{fe} mesh using arrays and transform them without for-loops. However, the default implementation of \jl{map} provided by Julia has some limitations related to memory requirements. The array $r$ returned by the \jl{map} function is an intrinsic Julia array by default, which implies that all entries $r_i$ are stored in memory for all indices $i$ simultaneously. It becomes a serious bottleneck when manipulating large arrays, e.g., cell-wise arrays for a \ac{fe} mesh with millions of cells.

Gridap provides the alternative function \jl{lazy_map}, which solves this memory bottleneck while exploiting the powerful high-level syntax provided by the \jl{map} function. Both \jl{map} and \jl{lazy_map} have the same syntax. The main benefit of using \jl{lazy_map} over the default \jl{map} function is that the former returns instances of a custom array type called \jl{LazyArray} instead of a built-in Julia \jl{Array}; see a simplified implementation of \jl{lazy_map} and \jl{LazyArray} in \lst{lst:lazy_array}. \jl{LazyArray} does not store the result $r_i$ for all indices $i$ simultaneously. Instead, it stores references to function $f$  and the input arrays $a^1,\ldots,a^n$, as shown in \lst{lst:lazy_array}. With this information, it computes on-the-fly the value $r_i$ when the index $i$ is requested by overloading the \jl{Base.getindex} function. 
We say that this array type is \emph{lazy} because it delays the computation of $r_i$ till its consumption. One potential drawback of this approach is that $r_i$ must be computed each time one gets a given index. 
 This approach is better than pre-computing all entries of large arrays in terms of memory requirements and allocations (see \sect{sec:efficient-elemental-operations} for more details about the latter). In addition, fetching pre-computed value from main memory comes with a cost and computing the value on-demand can be faster for some operations. Anyway, one can still use the default Julia \jl{map} when it can potentially be faster than \jl{lazy_map}, e.g., for small arrays. In Gridap, \jl{lazy_map} is used with arrays of length proportional to the number of the cells in the mesh. \jl{map} is used otherwise, e.g., in operations at the reference cell level. {Further optimizations can be considered when array entries are going to be accessed multiple times when positioned in a cell but are not covered here due to length restrictions.}

\begin{listing}[ht!]
\juliacode{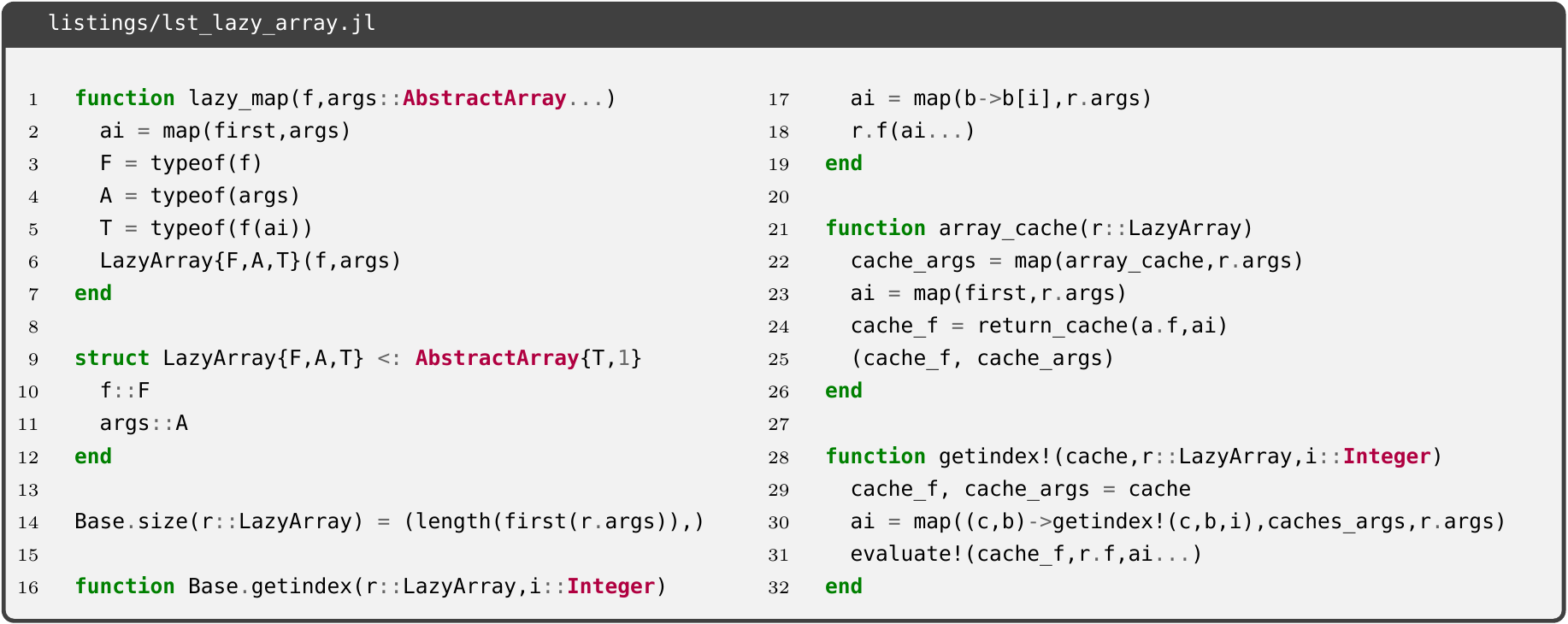}
\caption{Simplified implementation of \jl{lazy_map} and \jl{LazyArray}.
}
\label{lst:lazy_array}
\end{listing}

\begin{figure}[ht!]
\centering

\begin{subfigure}[b]{0.2\textwidth}
\centering
\includegraphics[scale=1]{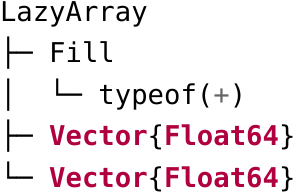}
\caption{}
\label{fig:lazy_op_tree-1}
\end{subfigure}
\begin{subfigure}[b]{0.2\textwidth}
\centering
\begin{tikzpicture}[sibling distance=6em]
  \node[nodest] (C) {\jll{c}}
    child { node[leafst,fill=mygreen!0] {\jll{a}} }
    child { node[leafst,fill=myblue!0] {\jll{b}} };
  \node[below= 0.2em of C.south] {\jll{+}};
\end{tikzpicture}
\caption{}
\label{fig:lazy_op_tree-2}
\end{subfigure}
\begin{subfigure}[b]{0.24\textwidth}
\centering
\includegraphics[scale=1]{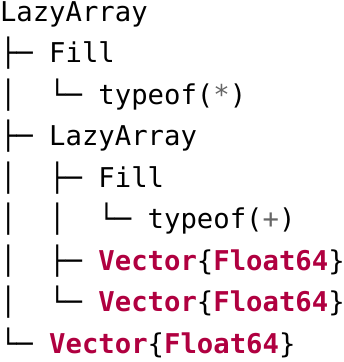}
\caption{}
\label{fig:lazy_op_tree-3}
\end{subfigure}
\begin{subfigure}[b]{0.24\textwidth}
\centering
\begin{tikzpicture}[sibling distance=6em]
  \node[nodest] (D) {\jll{d}}
    child { node[nodest] (C) {\jll{c}}
      child { node[leafst,fill=mygreen!0] {\jll{a}} }
      child { node[leafst,fill=myblue!0] {\jll{b}} }}
    child { node[leafst,fill=mygreen!0] {\jll{a}}};
    
  \node[below= 0.2em of C.south] {\jll{+}};
  \node[below= 0.2em of D.south] {\jll{*}};
\end{tikzpicture}
\caption{}
\label{fig:lazy_op_tree-4}
\end{subfigure}

\caption{Output of function \jl{print_op_tree} and graphical representation of the corresponding operation trees for arrays \jl{c} and \jl{d} in \lst{lst:lazy_op_tree}.}
\label{fig:lazy_op_tree}
\end{figure}
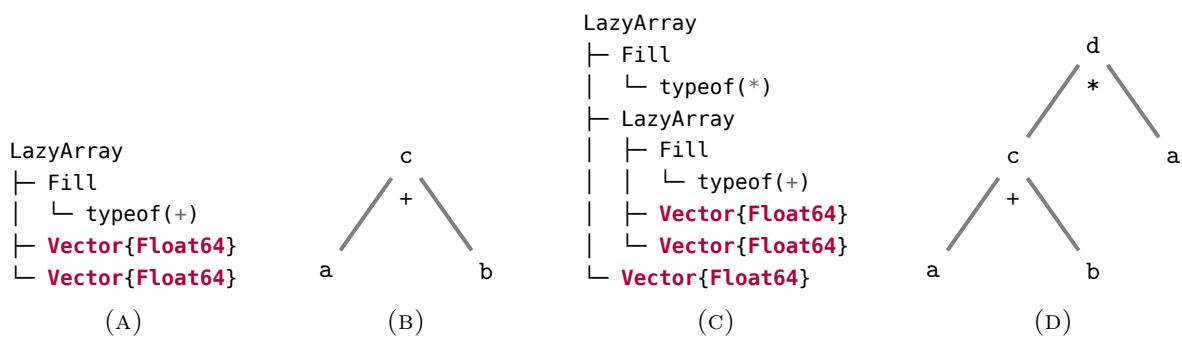

To illustrate the usage of \jl{lazy_map}, we consider the example in \lst{lst:lazy_op_tree}. Here, \jl{a} and \jl{b} are two standard Julia arrays of random values. We generate a new array \jl{c} by summing the entries of arrays \jl{a} and \jl{b} via the \jl{lazy_map} function. Array \jl{c} is an instance of the \jl{LazyArray} type. It contains a reference to the operation \jl{+} and the input arrays \jl{a} and \jl{b}. This information can be represented graphically as a small operation tree (see \fig{fig:lazy_op_tree-2}).  The operation tree can be visualized using the helper function \jl{print_op_tree}, which prints a representation of the tree in plain text format into the standard output. See the output of this function on array \jl{c} in \fig{fig:lazy_op_tree-1}.

From the output, it is clear that \jl{c} is of type \jl{LazyArray}. It has two inputs of type \jl{Vector{Float64}} combined via the operator \jl{+}.   A new lazy array \jl{d} with a more complex operation tree is generated by multiplying the entries of the lazy array \jl{c} with the entries of the array \jl{a}. In this case, \jl{d} stores a reference to operation \jl{*} and a reference to arrays \jl{c} and \jl{b}. Thus, the operation three associated with \jl{d} contains part of the operation tree of array \jl{c}, see \fig{fig:lazy_op_tree-3} and \fig{fig:lazy_op_tree-4}. It illustrates how one can create arbitrarily complex \jl{LazyArray} objects by nesting calls to the \jl{lazy_map} function. In some sense, the type \jl{LazyArray} is related with  types encoding symbolic operation trees in \ac{fe} codes based on symbolic \acp{dsl}. However, \jl{LazyArray} is not just a \emph{symbolic} type, but also a \emph{numeric} one, since it can be indexed to recover the numerical values represented by the array. This ambivalent nature of \jl{LazyArray} is one of the most powerful features of Gridap. It allows one to implement a high-level user API that resembles symbolic \ac{fe} frameworks and, at the same time, the objects involved have a numerical value that can be directly accessed without calling a C/C++ code generator.

\begin{listing}[ht!]
\juliacode{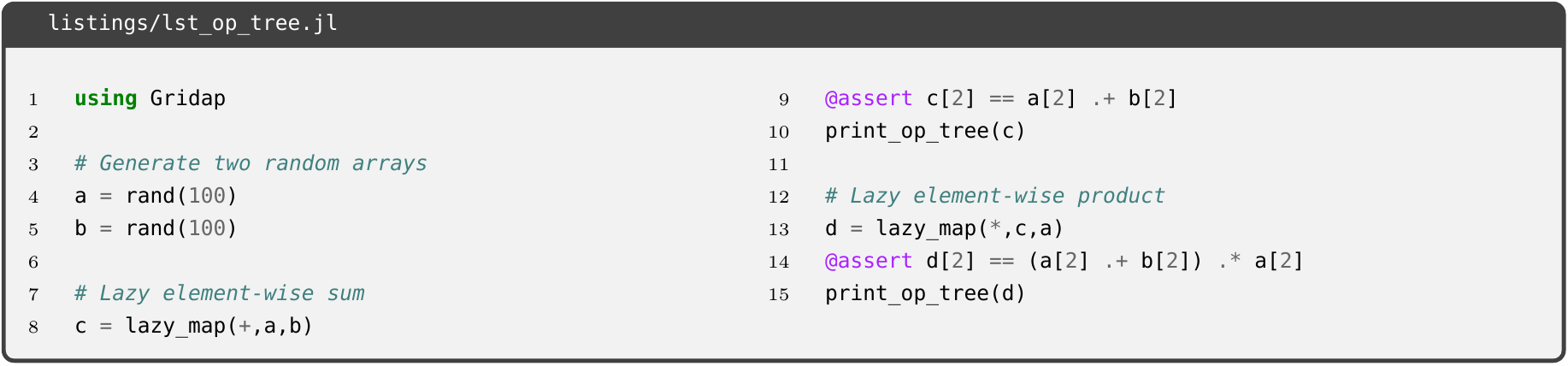}
\caption{Basic usage of the \jl{lazy_map} function provided by Gridap.}
\label{lst:lazy_op_tree}
\end{listing}

\subsection{Efficient elemental operations}\label{sec:efficient-elemental-operations} 
{Let us take a deeper look into \jl{lazy_map}. The performance of \jl{lazy_map} is compromised when the output or intermediate computations of $f$ require dynamic allocations. Elemental operations in \jl{lazy_map} can potentially be called many times and should be performant.} Gridap introduces the abstract type \jl{Map} to solve this problem. \jl{Map} represents functions or any other callable object able to pre-compute and reuse a cache to avoid dynamic allocation each time they are called. Specializations of the \jl{Map} type have to overload two new functions, namely \jl{return_cache} and \jl{evaluate!}, which generate the cache and evaluate the object by reusing this cache, respectively. 

Let us consider the example in \lst{lst:map} to illustrate the usage of these two functions. We start by creating two random Julia arrays \jl{a} and \jl{b}. Then, we create \jl{f} (an instance of the \jl{Broadcasting} type,  a specialization of \jl{Map}) that represents a \emph{broadcast} operation over some input arguments. Like in Python, a broadcast operation in Julia is an operation that is performed element-wise on some input arrays with implicit expansion of singleton axes (see the Julia documentation of further details). Calling an instance of this type is equivalent to broadcast the given operation on the input arrays. However, the main difference is that type \jl{Broadcasting} implements the interface of the abstract type \jl{Map} and thus, one can use function \jl{return_cache} and \jl{evaluate!} to pre-allocate and reuse the array resulting from the broadcast operation. \lst{lst:map} shows the basic usage of these functions. Function \jl{return_cache} computes a cache object containing a pre-allocated array storing the result of the broadcast operation. Then, \jl{evaluate!} uses this cache to compute the broadcast operation in-place without allocating the result.

\begin{listing}[ht!]
\juliacode{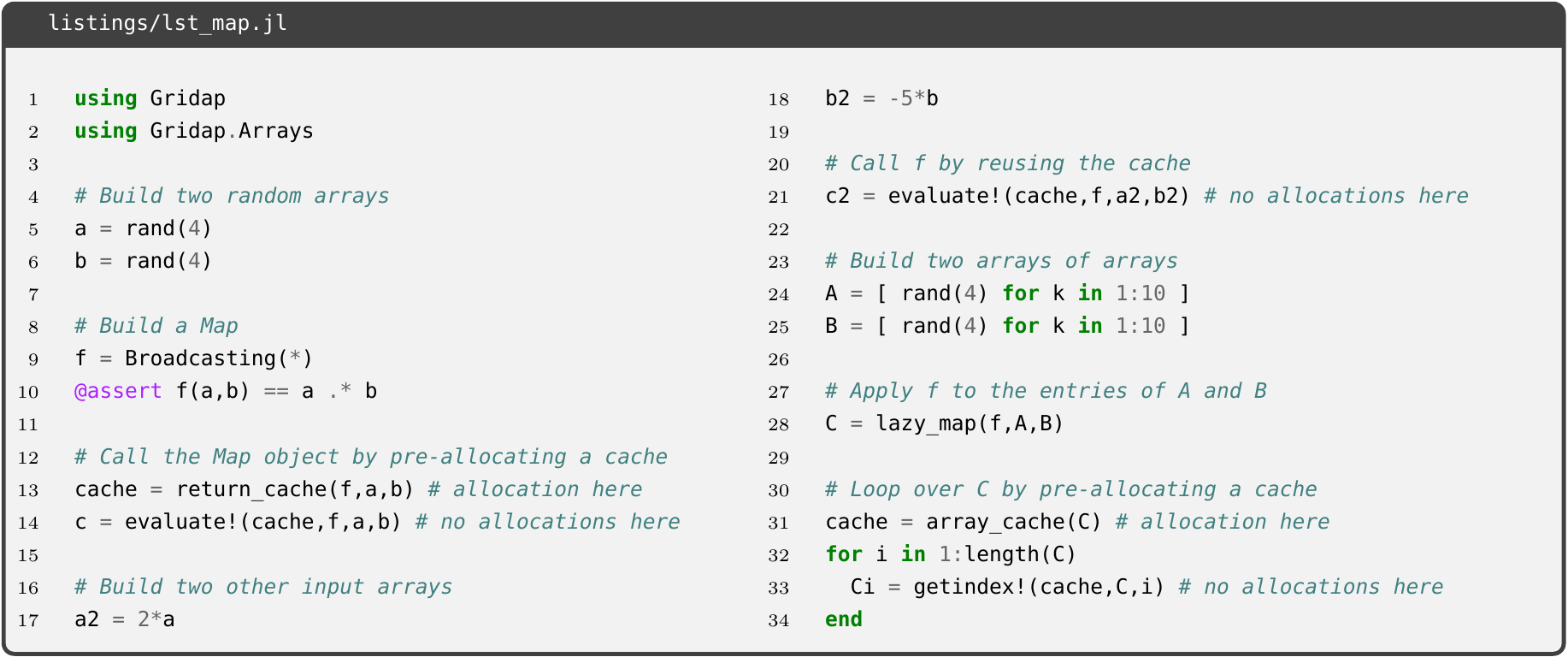}
\caption{Usage example of functions \jl{return_cache} and \jl{evaluate!}.}
\label{lst:map}
\end{listing}


One of the main applications of functions \jl{return_cache} and \jl{evaluate!} is to efficiently implement the \jl{LazyArray} type. By using these methods, it is possible to iterate an instance of \jl{LazyArray} without allocating memory at each step (see, e.g., the last loop in \lst{lst:map}). The implementation of \jl{array_cache} for the \jl{LazyArray} type (see \lst{lst:lazy_array}) builds all the cache objects needed to iterate an instance of \jl{LazyArray}, i.e., the cache objects needed to index the input arrays, and the cache needed to evaluate the elemental operation $f$. Then, the specialization of \jl{getindex!} for the \jl{LazyArray} type (see also \lst{lst:lazy_array}) uses these cache objects to compute a particular element of the array without allocating memory. See how the input arrays are indexed by using their cache objects and how the elemental operation $f$ is called by using its cache within the definition of \jl{getindex!} in \lst{lst:lazy_array}. In summary, the \jl{LazyArray} type does not only avoid allocating large arrays when mapping elemental operations on input arrays but also allows one to efficiently iterate on the result by enabling to reuse data at each iteration. This feature is at the heart of the efficient \ac{fe} implementation in Gridap.

\subsection{Representing fields} 
The actual implementation of \ac{fe} methods in Gridap starts with the definition of an interface representing fields. In Gridap, fields are understood as functions $f:D\subset\mathbb{R}^n\longrightarrow R$ that take vectors in a physical domain $D \subset \mathbb{R}^n$, with $0 < n \leq d$, and return values in a range $R$.
 In this context, we usually refer to elements of $D \in\mathbb{R}^n$ as \emph{evaluation points} since these are the points in the physical domain where fields are evaluated. Values in $R$ can be real numbers $\mathbb{R}$, complex numbers $\mathbb{C}$ or arbitrary order tensors with real or complex components, e.g., $R\subset\mathbb{R}^d$. In the code, the value of a field can be represented with any type specializing the abstract type \jl{Number} defined in Julia. Note that even though vector and tensor values are not numbers in a strict mathematical sense, we implement them as sub-types of \jl{Number} for convenience.  In particular, one can use the Julia types \jl{Float64} and \jl{ComplexF64} to represent the value of real-valued and complex-valued fields respectively, or the Gridap types \jl{VectorValue} and \jl{TensorValue} for vector-valued or second-order tensor-valued fields respectively (further details on the implementation of these tensor types will be provided later in \sect{sec:tensorvalues}). One can even consider more sophisticated number types, like \emph{dual numbers} to forward-mode automatic differentiation \cite{Khan2015} to automatically differentiate Gridap solvers in forward-mode \cite{Khan2015} using Julia packages like ForwardDiff~\cite{Revels2016}.

Different types of fields involved in a \ac{fe} computation, such as shape functions, \ac{fe} interpolations or manufactured solutions, are implemented as specializations of the abstract type \jl{Field}. The API of this type is a concretization of the API of the abstract type \jl{Map}. The \jl{Map}  type represents a general transformation between any input and output arguments, whereas \jl{Field} stands for the particular case in which the input is a point in $\mathbb{R}^n$ (an instance of \jl{VectorValue}) and the output is a number or a vector/tensor (an instance of some specialization of \jl{Number} in general). The abstract type \jl{Field} is defined as a  sub-type of \jl{Map} and, in consequence, all the generic code implemented for \jl{Map} objects can also be used with \jl{Field} instances. 

In order to implement a new \jl{Field} type one needs to specialize functions \jl{return_cache} and \jl{evaluate!}, just as previously explained for \jl{Map}. With these methods, it is possible to use a set of generic methods implemented for \jl{Field} such as algebraic operations, function composition, differential operators, etc. (See \sect{sec:field_ops} for further details). In order to illustrate how a new field type is implemented, \lst{lst:fields} contains a simplified implementation  of the Gridap type \jl{ConstantField} that represents a field with a  constant value, i.e. $f(x)$ is defined in this case as $f(x)\doteq\hat f$ for some number $\hat f$ and for any evaluation point $x$. This specialization optimizes several computations in the code. E.g., the Jacobian of the geometrical map in linear simplicial meshes is a \jl{ConstantField} in Gridap, which makes the evaluation of the Jacobian at the integration points very efficient. \jl{ConstantField} is a specialization of \jl{Field}  that stores the value of the field in the variable \jl{value}, returned each time the field is evaluated. The last argument in the definition of \jl{evaluate!} is the point \jl{x} where the field is to be evaluated. Object \jl{x} is an instance of \jl{Point}, which is an alias of \jl{VectorValue} used to emphasize that vector \jl{x} is interpreted as an evaluation point in this context. The implementation of this function is trivial for this particular case since it just needs to return the value stored within the given instance of \jl{ConstantField}. However, other field specializations will consider the evaluation point to compute the result. Note also that the cache object is not used in this simple example since the result is already pre-computed, but other field implementations can also overload function \jl{return_cache} and make use of the cache object if needed as explained before for the \jl{Map} abstract type. Once the \jl{evaluate!} function is defined, one can, e.g., evaluate the resulting field objects using function notation or evaluating the field at an array of points, as demonstrated in the last part of \lst{lst:fields}. In the latter case, the result is a Julia vector containing the field values at all points that needs to be heap-allocated. Thus, it is recommended to generate a cache object and then evaluate the field without memory allocations. In order to implement a new \jl{Field} type one needs to specialize functions \jl{return_cache} and \jl{evaluate!}, just as previously explained for \jl{Map}. Once these methods are implemented, it is possible to use a set of generic methods implemented for \jl{Field} such as algebraic operations, function composition, differential operators, etc (see \sect{sec:field_ops} for further details). In order to illustrate how a new field type is implemented, \lst{lst:fields} contains a simplified implementation  of the Gridap type \jl{ConstantField} that represents a field with a  constant value, i.e. $f(x)$ is defined in this case as $f(x)\doteq\hat f$ for some number $\hat f$ and for any evaluation point $x$. This specialization is used to optimize several computations in the code, e.g., the Jacobian of the geometrical map in meshes of linear simplices is represented in Gridap with a \jl{ConstantField}, which makes the evaluation of the Jacobian at the integration points very efficient. The struct \jl{ConstantField} is defined as a specialization of \jl{Field}  and it stores the value of the field in the variable \jl{value}, which can be returned each time the field is evaluated. In the definition of function \jl{evaluate!}, the last argument is the point \jl{x}, where the field is to be evaluated. Object \jl{x} is an instance of \jl{Point}, which is an alias of \jl{VectorValue} used to emphasize that vector \jl{x} is interpreted as an evaluation point in this context. The implementation of this function is trivial for this particular case, since it just needs to return the value stored within the given instance of \jl{ConstantField}. However, other field specializations will consider the evaluation point to compute the result. Note also that the cache object is not used in this simple example since the result is already pre-computed, but other field implementations can also overload function \jl{return_cache} and make use of the cache object if needed as explained before for the \jl{Map} abstract type. Once the \jl{evaluate!} function is defined, one can e.g., evaluate the resulting field objects using function notation or evaluating the field at an array of points, as demonstrated in the last part of \lst{lst:fields}. In the latter case, the result is a Julia vector containing the value of the field at all evaluation points. Since the result is a vector that needs to be heap-allocated, it is recommended to generate a cache object and then evaluating the field at a given vector of points as many times as needed without allocating extra memory.

\begin{listing}[ht!]
\juliacode{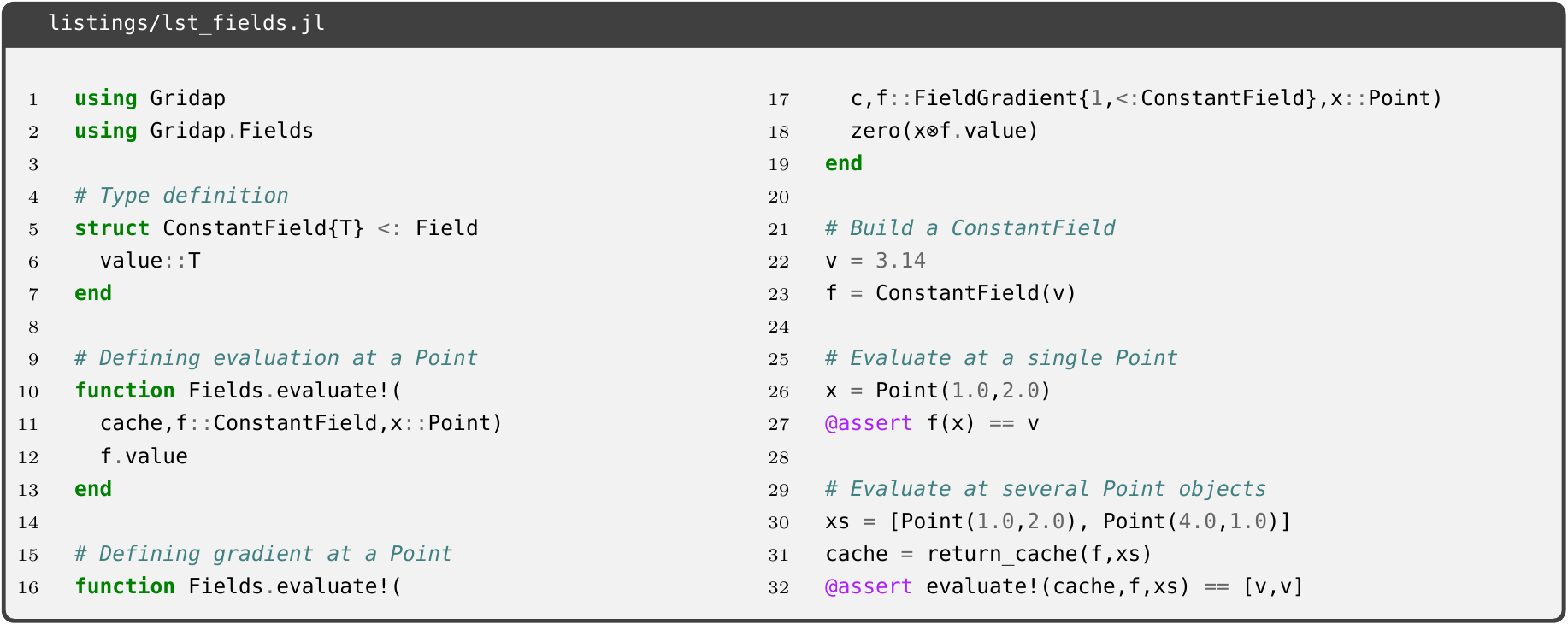}
\caption{Implementation and usage of a new field type in Gridap.}
\label{lst:fields}
\end{listing}

\subsection{Tensor values and efficient tensor algebra}
\label{sec:tensorvalues}

Gridap provides a family of concrete types representing tensors of different orders. First order tensors (i.e., vectors) are represented with instances of the type  \jl{VectorValue{D,T}}, where the type parameters  \jl{D} and \jl{T} are the vector length and component type respectively. Similarly, second-order tensors are represented with instances of the type \jl{TensorValue{D1,D2,T}}, where the size of the tensor is \jl{D1} times \jl{D2} and the component type is \jl{T}. Other related types include symmetric second-order tensors (e.g., stress and strain in mechanical computations), third-order tensors (e.g., the gradient of vector-valued fields) and symmetric fourth-order tensors (e.g., constitutive relations), but they are not detailed here for the sake of brevity. All these concrete tensor types inherit from the abstract type \jl{MultiValue}, which serves as the umbrella that covers all these cases. Gridap defines several tensor algebra operations for tensor-values, like scaling, addition, subtraction, dot product, single and double contractions, inner products, and outer products. As an example, the first part of \lst{lst:tensors} shows how to perform some operations with first and second-order tensors.\footnote{LaTeX symbols can be easily written in the Julia REPL by typing the corresponding LaTeX macro and pressing the tabulator. E.g., \jl{α} is written as \jl{\alpha} + \jl{Tab}. Several text editors and IDEs (like Vim, Emacs and VSCode) provide Julia extensions to write latex symbols this way.} 

The implementation of tensor operations as the ones showed in \lst{lst:tensors} is one of the most performance-critical parts in a \ac{fe} code since one needs to use them with very low granularity, i.e., at each integration point and for each local \ac{dof} at each cell of the \ac{fe} mesh. For this reason, Gridap leverages many advanced Julia features to achieve an efficient implementation. An important design decision in this regard is to include the size of the tensors in the type parameters of the corresponding types. By doing so, the size of the tensors is known at \emph{compile time} instead of at \emph{run time}, which makes it possible to introduce some important optimizations. In particular, it allows one to specialize operations for particular tensor sizes and unroll for-loops to achieve performance.  For instance, the dot product of two instances of \jl{VectorValue{2,T}} (i.e., vectors of 2 components) can be implemented efficiently for this particular size by explicitly writing the result {by loop unrolling} (see line 20 of \lst{lst:tensors}). Even though the number of different tensor sizes in a \ac{fe} simulation is bounded (the number of space dimensions is usually $d\leq 4$), writing all specializations in this way can be tedious. To solve this problem, Julia provides an advanced feature called \emph{generated} functions, which are functions with a special syntax. In a conventional Julia function, the function body includes operations that compute and return a value at run-time. In contrast, the function body of a generated function contains commands to generate and return a piece of code that, once compiled, computes and returns the value of the function at run time. A generated function can be interpreted as a  pre-compilation step in which the code to be compiled is generated using information about the types of the input arguments.  For instance, in line 26 of \lst{lst:tensors} we show how to use a generated function to define the dot product of two instances of \jl{VectorValue{D,T}} for a generic \jl{D}. The body of the function generates the code to be compiled for a particular \jl{D}. Here, this is done by generating a string containing the code and parsing it, which returns an \jl{Expr} object that is the format used in Julia to represent Julia code. Note that the routines that generate the code are invoked once, right before the function compilation, whereas the routines within the generated code are called in run-time. As a result, the for-loop that generates the code is executed once in the compilation stage and the code that is called at run-time has no for-loops, thus achieving performance.

\begin{listing}[ht!]
\juliacode{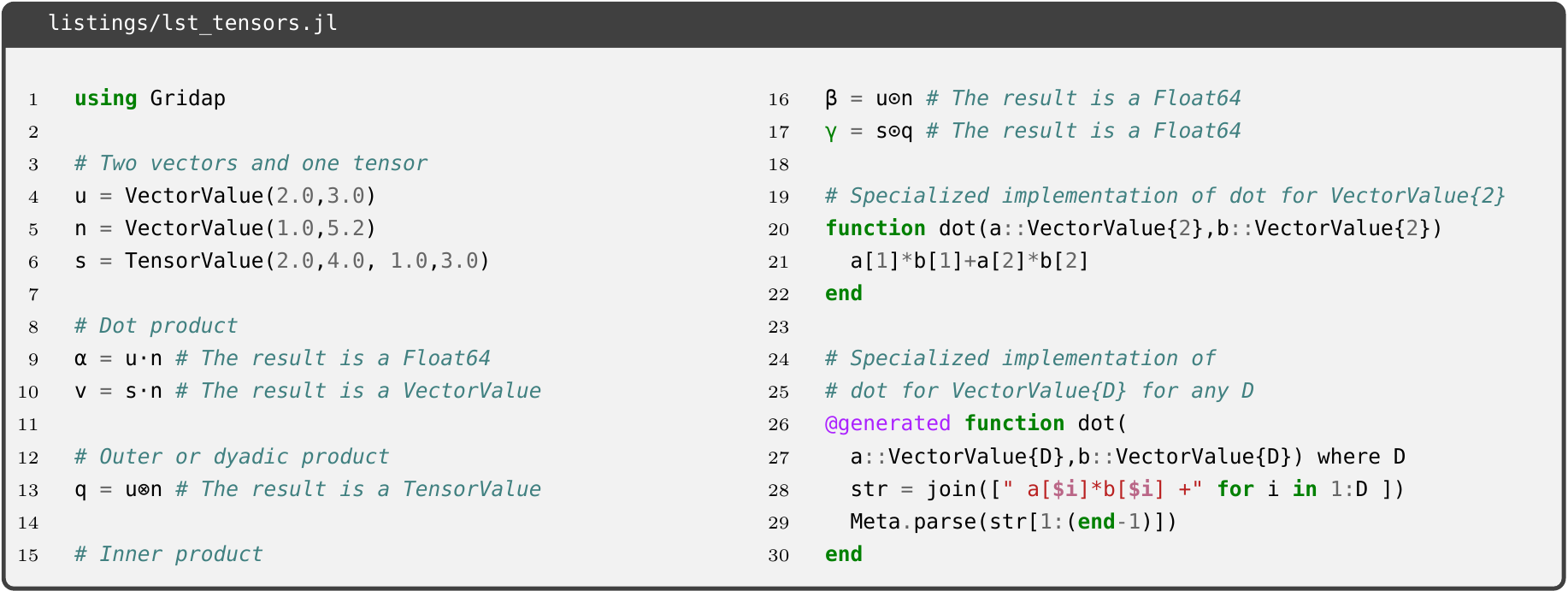}
\caption{Overview of tensor operations available in Gridap and their efficient implementation.}
\label{lst:tensors}
\end{listing}

\subsection{Algebraic and differential operations on fields}
\label{sec:field_ops}

Gridap does not only allow one to operate on values but also to operate on fields directly.  Let us assume that $\diamond$ denotes some operation, e.g., $+$ or any other binary operation over fields $f$ and $g$ defined on the same domain. In Gridap, it is possible to create a new field $f\diamond g$, that represents the result of applying operation $\diamond$ to $f$ and $g$, namely $(f\diamond g)(x)\doteq f(x) \diamond g(x)$. The result field is an instance of a concrete field type called \jl{OperationField}, which contains references to the given operation and the input fields. When an \jl{OperationField} is evaluated, e.g., with \jl{evaluate!}, the input fields are evaluated first. Next, the operation is applied to the result. One of the main advantages of operating over fields (instead of evaluating the fields first and then operating over the values) is that the result is an instance of a sub-type of \jl{Field}. Thus, one can use it in any generic function implemented for \jl{Field} objects. This feature is heavily used in the library, e.g., in functions performing numerical integration or visualization implemented in terms of the abstract type \jl{Field}, thus reducing code duplication.
  
 \begin{listing}[ht!]
\juliacode{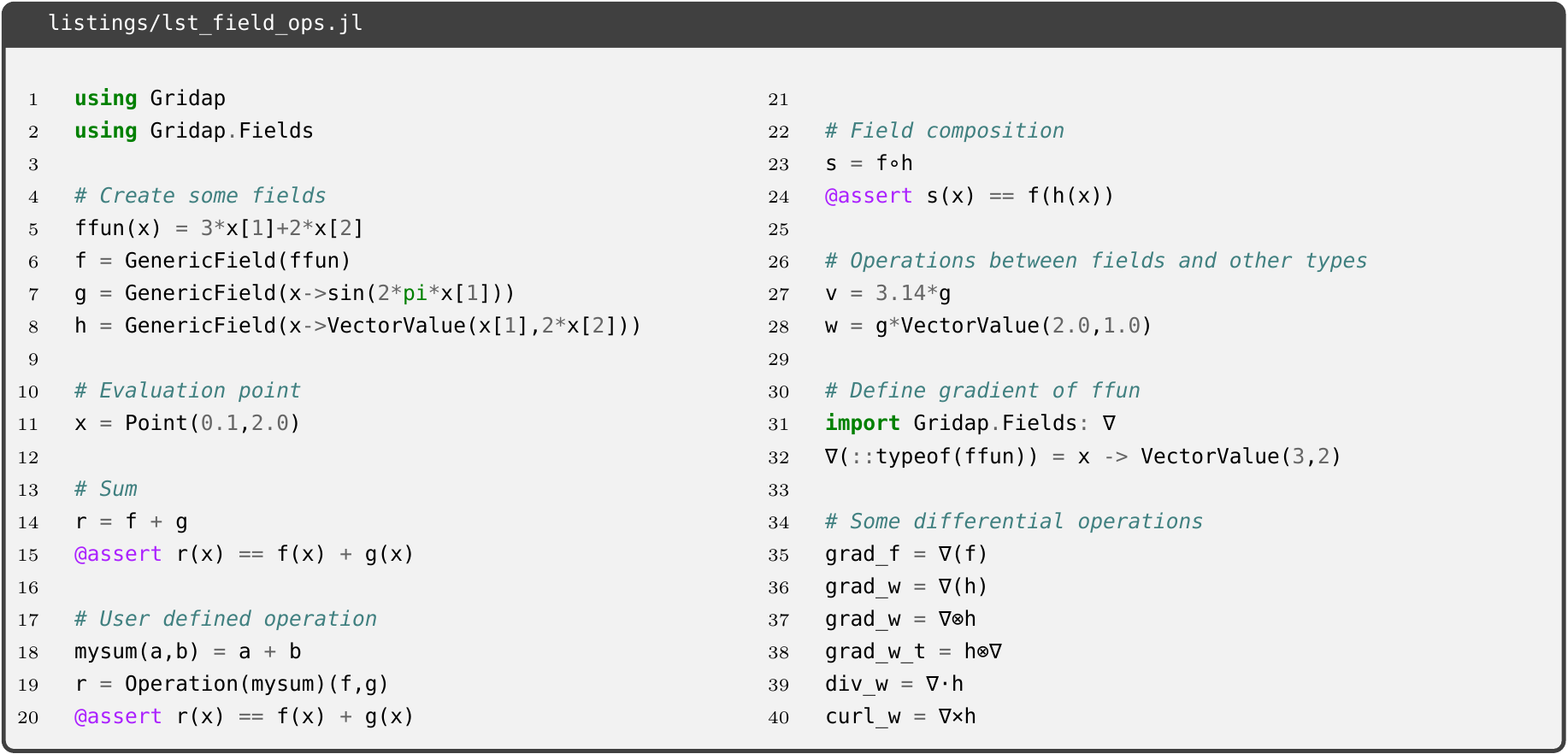}
\caption{Overview of algebraic and differential operations on fields.}
\label{lst:field_ops}
\end{listing}

The syntax for operating fields is illustrated in \lst{lst:field_ops}. We start the code snippet by creating three instances of \jl{GenericField}. This type is a specialization of the abstract \jl{Field} type used to generate fields from objects that behave like fields, but their type is not a specialization of \jl{Field}. A typical application is to build fields from analytical functions, as shown in this piece of code. Note that the generated fields, \jl{f} and \jl{g} are scalar-valued, whereas \jl{h} is vector-valued (we specifically need a vector-valued field in some of the operations displayed in the example). For basic operations defined in Julia like \jl{+} one can write operations on fields as they were operations on values. However, user-defined operations must follow an alternative syntax because they cannot be anticipated. In this case, one uses the auxiliary constructor \jl{Operation} that returns an object that can be evaluated to perform the desired operation over the input fields.  In addition, the constructor \jl{Operation} allows us to implement the function composition $f\circ h$ of two fields $f$ and $h$, where the range of $h$ is equal to the domain of $f$, namely $(f\circ h)(x) = f(h(x))$. The call \jl{f∘h} is simply implemented as \jl{Operation(f)(h)} since \jl{f} can be interpreted as a user-defined unary operation on the (vector) value of \jl{h}. It is also possible to operate fields with instances of other types like numbers or functions. In this case, the objects that are not fields are converted into fields internally before doing the operation. Numbers are converted into \jl{ConstantField} objects and functions into \jl{GenericField} objects. In addition to algebraic operations, Gridap also defines several differential operators like gradient, divergence, and curl. See last lines in \lst{lst:field_ops} for some particular examples.  The result of a differential operation on a \jl{Field} object is also a \jl{Field} object that can be used in other parts of the code. When differentiating fields defined from user-defined functions, like the fields \jl{f} and \jl{h} in \lst{lst:field_ops}, Gridap considers the automatic differentiation package ForwardDiff \cite{Revels2016}  to automatically compute the derivatives under-the-hood. However, it is also possible to bypass the automatic differentiation step if needed by explicitly defining the gradient, as shown in line 32 of \lst{lst:field_ops}.

\subsection{Representing linear functionals} 
The final low-level ingredient in the computational back-end is the abstract type \jl{Dof} that represents a linear functional on some space of fields. The name of this type is motivated by the Ciarlet definition of a \ac{fe} element \cite{Ciarlet2002}, where the linear functionals defining the shape functions are referred to as \acp{dof}. The abstract type \jl{Dof} represents callable objects that take a  \jl{Field} object and return a real number via a linear transformation. The abstract type \jl{Dof} inherits from \jl{Map} and, thus, a \jl{Dof} object can be used in any generic function implemented for \jl{Map}.  All the functionality previously presented for the abstract types \jl{AbstractArray}, \jl{Field} and \jl{Dof} is modular and composable and can be combined to construct a myriad of different objects involved in a \ac{fe} computation. To illustrate this, let us consider the following two practical examples.

\subsection{Example: reference shape functions} 

In this example, we demonstrate a possible way to build the shape functions of a  3-node reference triangle via the low-level API of Gridap and how to use these functions to interpolate some given nodal values. To this end, we start from a monomial basis spanning the 1st order ($k=1$) 2-variate polynomial space P on the vector variable $x$, namely $M\doteq\{x^{e_1}_1x^{e_2}_2:\ e_1+e_2\leq k\}$. From these monomials, we obtain the shape functions with the change of basis $\hat s_i \doteq \sum_j A^{-1}_{ji} m_j$, where $\hat s_i$ is the $i$-th shape function, $m_j$ is the $j$-th monomial in $M$, and $A$ is the inverse of the change of basis matrix. We compute $A_{ij}\doteq \hat l_i(m_j)$ by introducing a basis $\{\hat l_i:\ i=1,\ldots|M|\}$ of the dual space of $M$, i.e., the basis of \acp{dof}. In this case, we consider the Lagrangian \ac{dof} basis associated with the 3-nodded reference triangle and $\hat l_i(m_j)$ is simply the evaluation of the monomial $m_j$ at the coordinate vector of the $i$-th node in the reference triangle. We build Lagrangian shape functions for simplicity, but other \ac{fe} spaces (like Raviart-Thomas or Nédélec) can be constructed similarly.

\begin{listing}[ht!]
\juliacode{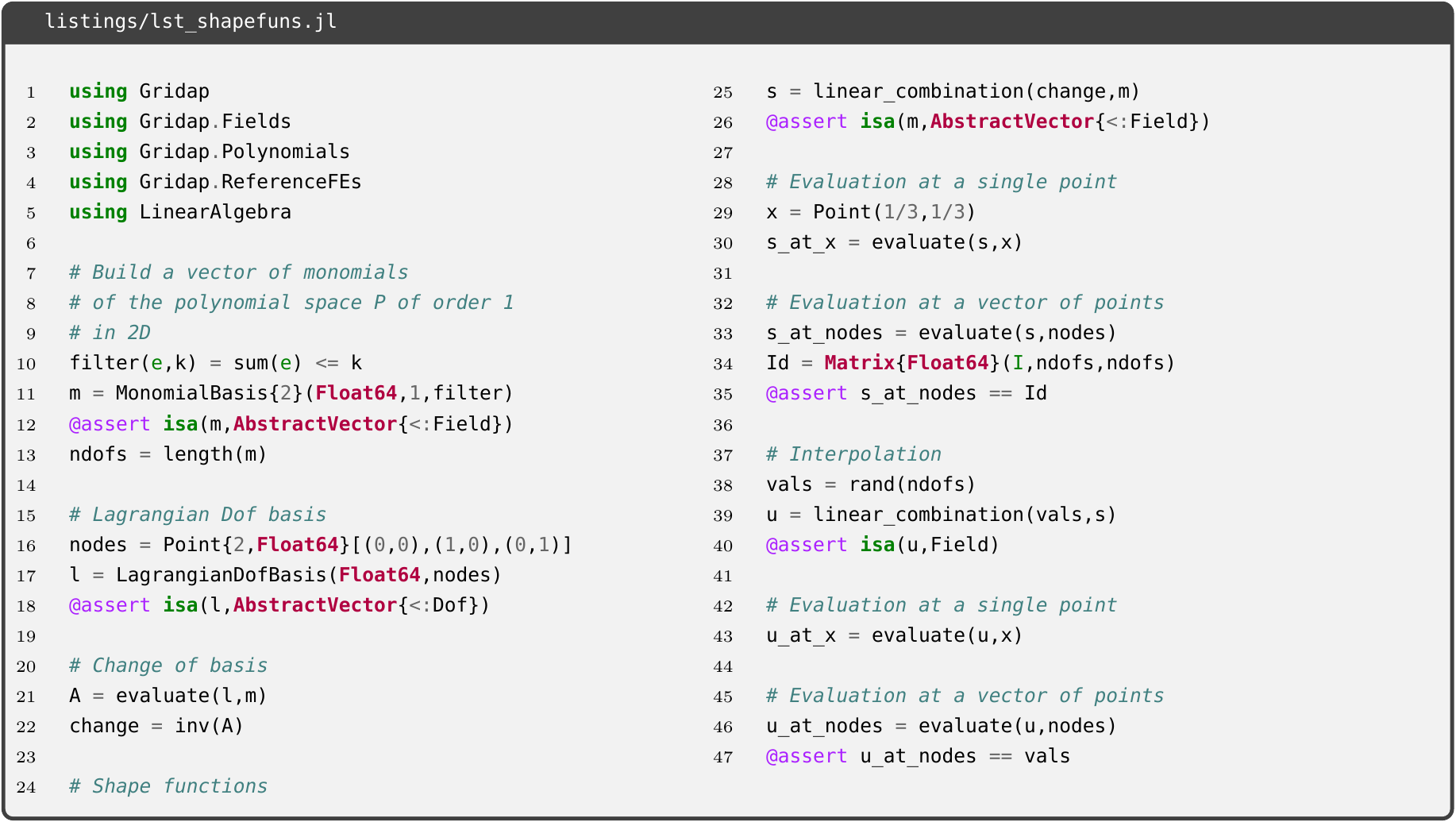}
\caption{Construction of the shape functions for a 1st order P element in 2D and interpolation of some user-defined nodal values via the low-level API of Gridap.}
\label{lst:shapefuns}
\end{listing}

The actual computation of the shape functions is given in \lst{lst:shapefuns}. We start by building the monomial basis via the constructor \jl{MonomialBasis}, which accepts several inputs. In this case, we chose 1st order 2-variate monomials with values represented by a \jl{Float64} number.  The function object passed in the last argument is a filter to select the subset of the tensor-product multivariate space Q to be included. We define a function \jl{filter} to select only monomials in the P space; it returns \jl{true} for $e_1+e_2\leq k$ and false otherwise. Note that the inputs \jl{e} and \jl{k} of function \jl{filter} correspond to the tuple $(e_1,e_2)$ and the value $k$ respectively.  The resulting object \jl{m}, containing the monomial basis, is a vector of \jl{Field} objects (as checked after its construction) with a field per monomial in the basis. This object can be consumed by the methods defined in the abstract interfaces of \jl{AbstractArray} and \jl{Field} generically.  The subsequent lines compute the change of basis that transform the monomials into the shape functions. To this end, we build the Lagrangian \ac{dof} basis associated with the nodes of the reference triangle. The resulting object \jl{l} is as a vector of \jl{Dof} objects. Each one of those \acp{dof} corresponds to the evaluation at a node in the reference triangle. Then, we compute matrix \jl{A} by evaluating the \ac{dof} basis \jl{l} on the monomials \jl{m}.  The shape functions are constructed with a linear combination of the coefficients in the inverse of matrix \jl{A} and the monomial basis \jl{m} (see, e.g., \cite{Badia2018} for more details). Note that function \jl{linear_combination(x,y)} is just a short-hand for \jl{transpose(x)*y}. The resulting value \jl{s} containing the shape functions is also a vector of \jl{Field} objects. It can be consumed using the generic methods associated with \jl{AbstractArray} and \jl{Field}, just like for the monomials. E.g., we evaluate the shape functions at the reference nodal coordinates to check that the resulting matrix is indeed the identity as expected. Note that shape functions (and any \jl{Field} object) can be evaluated at arbitrary points, not only nodes of the reference element or integration points. This is useful for interpolation purposes. E.g., high-order \ac{fe} fields can be visualized by interpolating them in linear spaces on finer grids. The final lines in the \lst{lst:shapefuns} show how one can interpolate some nodal values using the previously computed shape functions. The interpolation is simply built as a linear combination of the nodal values and the shape functions, namely $u(x)=\sum_i u_i \hat s_i(x)$, where the vector \jl{vals} stores the coefficients $u_i$. The result \jl{u} is a \jl{Field} object that can also be evaluated at arbitrary points. We check that \jl{u} is correct by evaluating it at the nodal coordinates and checking that the resulting vector coincides with the vector of nodal values. Note that the code giving support to the low-level API in this example works for any space dimensions. As a result, this code snippet can be easily modified to compute the shape functions of a simplex in any other number of spatial dimensions.

\subsection{Example: cell-wise elemental stiffness matrices}
\label{sec:cellmatlow}

In this second example, we go a step further and illustrate how one can use the low-level API of Gridap to build a vector containing the elemental stiffness matrix for each cell in a computational mesh. In this example, we use a simple 2$\times$2 Cartesian mesh of the domain $(0,2)^2$ as an easy way to initialize the vector of nodal coordinates and cell connectives of the mesh by hand. These values can also come from a mesh generated via a mesh generator (further details later in \sect{sec:frontend}). The goal of this example is to compute component $ij$ of the stiffness matrix $K^e$ for each cell id $e$ in the mesh, namely
\begin{equation}
[K^e]_{ij} \doteq \int_{\hat\Omega} (J^{-{\rm t}}_e\cdot \nabla \hat s_i)\cdot (J^{-{\rm t}}_e\cdot \nabla \hat s_j)\ |\det(J^{\rm t}_e)| \ {\rm d}\hat\Omega.
\end{equation}
Here, $\hat s_i$ is the $i$-th Lagrangian shape function in the reference quadrilateral $\hat\Omega\doteq (0,1)^2$, $J^{\rm t}_e\doteq \nabla \varphi_e$ is the transposed Jacobian, and $\varphi_e:\hat\Omega\longrightarrow \Omega_e$ is the isoparametric map,  $\varphi_e(x)\doteq \sum_i x^e_i \ \hat s_i(x)$, being $x^e_i$ the coordinate vector of the $i$-th local node in the $e$-th cell of the mesh. Note that the gradient operator $\nabla$ is the transpose of the derivative.

The code of this example is in \lst{lst:cell_mat}. The main goal of the code snippet is to build a vector \jl{cell_mat} that represents the elemental stiffness matrices for all cells of the mesh; the stiffness matrix for cell id \jl{e} is fetched as \jl{cell_mat[e]}. The code in \lst{lst:cell_mat} is intentionally very low-level to illustrate the fundamental building blocks of the library but an average user rarely needs to build all these quantities manually. (The high-level API is discussed in \sect{sec:frontend} and used in \lst{lst:cell_mat_user} to build \jl{cell_mat}.) 
We start by defining the computational mesh via the plain vector of nodal coordinates and the vector of cell connectivity. Then, we built the reference shape functions for this mesh, which is done via a small variation of the code previously discussed in \lst{lst:shapefuns}, since now we need to build Lagrangian shape functions for the reference quadrilateral instead of the reference triangle. Then, we define the vectors of integration points and weights to integrate quantities in the reference element. We use a single integration point (enough to integrate the elemental matrix exactly in this case), but one can use an arbitrary number. The next step is to build the cell-wise vector of reference shape functions \jl{cell_s}. This vector can be indexed by a cell id \jl{e}, namely \jl{cell_s[e]}, to get the reference shape functions for this particular cell. Note that all variables named \jl{cell_*} in the code snippet are indexed by a cell id, which is a common naming convention in Gridap. In this example, all the cells have the same reference element. Thus, we use a \jl{Fill} vector to represent \jl{cell_s} in a memory-efficient way. Internally, \jl{cell_s} stores a single reference to \jl{s} that is returned each time the vector is indexed instead of storing and returning an independent copy of \jl{s} for each cell. As a result, the memory consumed by \jl{cell_s}  is independent of the number of cells. We avoid using a standard Julia vector since it would consume an amount of memory proportional to the number of cells. We also use \jl{Fill} to build the cell-wise vectors of reference quadrature points and weights \jl{cell_q} and \jl{cell_w} respectively for the same reason.

\begin{listing}[ht!]
\juliacode{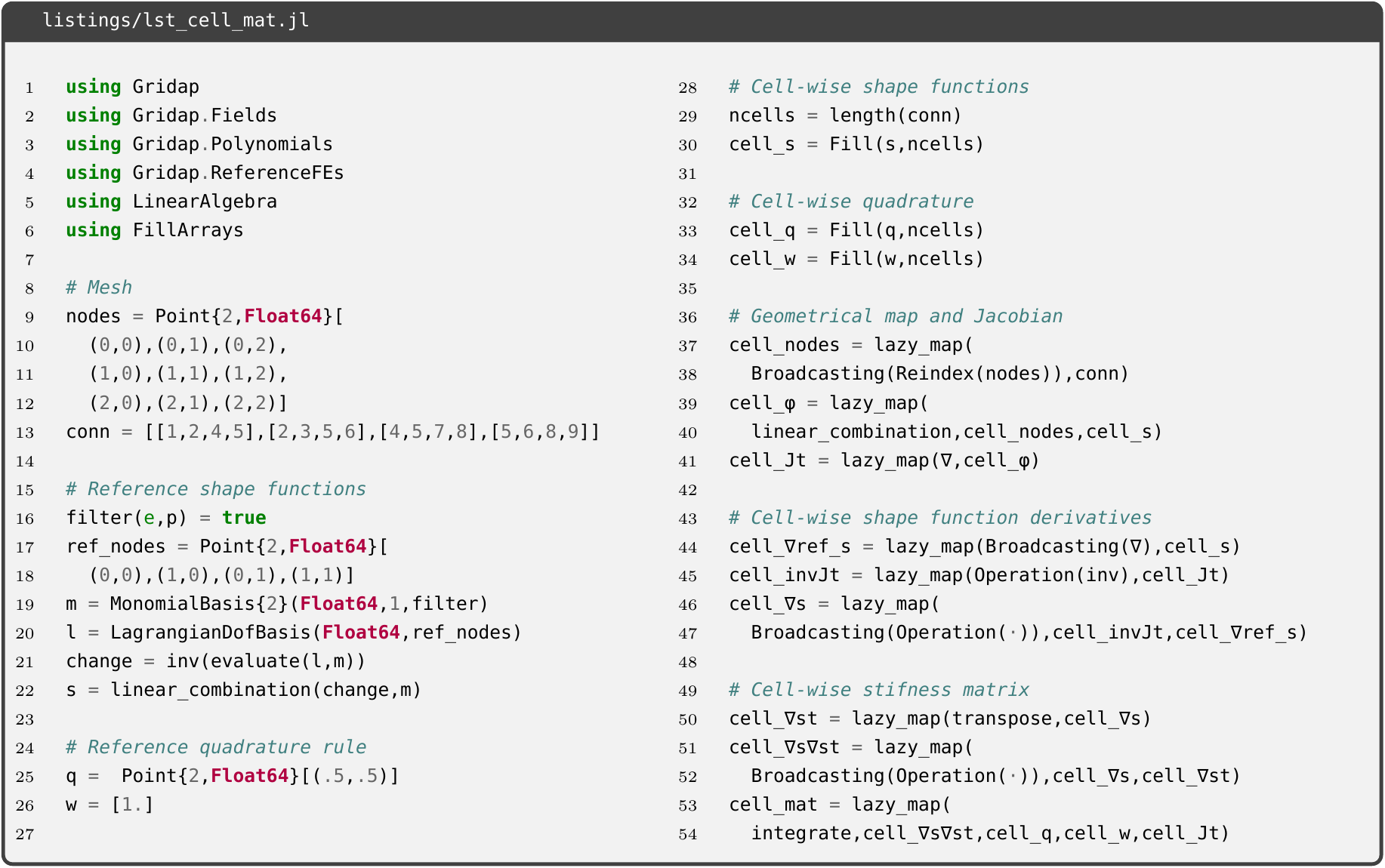}
\caption{Creating a \emph{lazy} vector of elemental stiffness matrices via the low-level API of Gridap.}
\label{lst:cell_mat}
\end{listing}

After this preliminary setup, we build the geometrical map and its Jacobian for all cells in the mesh. To this end, we need to build the cell-wise vector of nodal coordinates \jl{cell_nodes} such that \jl{cell_nodes[e]} returns the vector of \emph{physical} nodal coordinates for cell id \jl{e}. This vector is computed in terms of the nodal coordinates \jl{nodes} and the cell connectivity \jl{conn} using \jl{lazy_map} and the elemental operations \jl{Broadcasting} and \jl{Reindex}. The operator \jl{Reindex} takes an array and transforms it into a callable object \jl{f=Reindex(x)} such that \jl{f(i) = x[i]} for an index \jl{i}. This allows one to create the cell-wise nodal coordinates \jl{cell_nodes} by applying cell-by-cell the elemental operator \jl{Broadcasting(Reindex(nodes))} to the cell-wise node ids in \jl{conn}. Note that the memory consumed by the resulting array \jl{cell_nodes} is independent of the number of cells in the mesh since it is an instance of \jl{LazyArray}, which contains references to the input arrays \jl{conn} and \jl{nodes} instead of storing all the cell-wise nodal coordinates explicitly. At this point, one can build the cell-wise geometrical map with a cell-by-cell linear combination of the nodal coordinates in \jl{cell_nodes} and the shape functions in \jl{cell_s}. The resulting array \jl{cell_φ} is a lazy vector of \jl{Field} objects, representing the cell-wise geometrical map. On can further apply the gradient operator cell-by-cell to this array to end up with the transpose of the Jacobian at each cell. The next section of the code snippet deals with the computation of the gradient of the shape functions with respect to the physical coordinate with formula $J^{-{\rm t}}_e\cdot \nabla \hat s_i$. The resulting array \jl{cell_∇s} is a vector of vectors of \jl{Field} objects containing the \emph{physical} gradient of the shape functions for all the cells. The next step is to prepare the integrand of the elemental matrices, \jl{cell_∇s∇st}, which is built from \jl{cell_∇s} and its cell-wise transpose \jl{cell_∇st}. At a given cell id \jl{e}, \jl{cell_∇s[e]} is a column vector and \jl{cell_∇st[e]} is a row vector. Thus,  by broadcasting 
the dot product over a column vector of length say $n$ and a column vector of size $1\times n$ we end up with a matrix of size $n\times n$ representing the components of the stiffness matrix before integration at each one of the underlying cells. The cell-wise stiffness matrix \jl{cell_mat} is computed by integrating these objects cell-by-cell using the quadrature rule and the Jacobian of the geometrical map.

\section{The high-level API}
\label{sec:frontend}

\subsection{From low-level to high-level}

All the low-level machinery presented in previous sections is sufficient to implement the computation of elemental matrices and vectors for a wide range of weak forms. In any case, this machinery is general and outreaches \ac{fe} applications. However, these low-level functions require an in-depth knowledge of the library and are not convenient for non-expert users willing to use the library at a higher level. To solve this issue,  Gridap introduces higher-level objects that wrap the low-level arrays and related functions, providing a more convenient API that is safer and simpler to use in the majority of cases. A Gridap user rarely needs to deal with the low-level arrays directly. Only library developers and advanced users need to be aware of the low-level API.
{In contrast, the high-level API of Gridap described below is specific to \ac{fe} implementations.}
It relies on the abstract type \jl{CellDatum}, which represents a generic object containing an array describing a quantity defined on the cells of a mesh (i.e., each entry in the array is associated with a cell id), plus some meta-data that gives useful information not available in the plain array itself. This metadata includes the mesh on top of which the cell-wise array is defined and a \emph{trait} indicating whether the cell-wise data is defined in the \emph{reference} or \emph{physical} domain of the corresponding mesh. Concrete types implementing the \jl{CellDatum} interface need to implement three main functions, namely \jl{get_data}, \jl{get_triangulation}, and \jl{DomainStyle}. Function \jl{get_data} returns the plain array wrapped in the \jl{CellDatum} object, whereas the other two functions provide the extra meta-data. On the one hand, \jl{get_triangulation} returns the underlying mesh,  encoded via the abstract type \jl{Triangulation}.  There are two main types for representing computational grids in Gridap, namely \jl{DiscreteModel} and \jl{Triangulation}. The former contains all the data provided by mesh generators like Gmsh, namely node coordinates, cell and lower-dimensional objects (like vertices, edges and faces) connectivities, and face labels for imposing boundary conditions. On the other hand, the \jl{Triangulation} type represents a discretization of an integration domain. Note that one can extract different \jl{Triangulation} objects from the same \jl{DiscreteModel} since a \ac{fe} mesh can represent different integration domains (e.g., the cells in the bulk of the meshes and the faces on the boundary would define two different integration domains). On the other hand, \jl{DomainStyle} returns either \jl{ReferenceDomain()} or \jl{PhysicalDomain()} indicating whether the underlying data is defined in the reference or the physical domain respectively. This trait allows the user to handle high-level objects without knowing whether the underlying data is in the reference or physical domain. It allows us to accommodate a wide spectrum of \ac{fe} methods in the same framework since the library is not limited to interpolations defined in the reference space only.  In summary, working with \jl{CellDatum} objects is more convenient than working with the plain array directly because the stored meta-data allows us to implement safety checks and implicit conversions when manipulating the object.

To illustrate this, let us consider the code in \lst{lst:cell_datum}.  In this snippet, we build a \jl{CellField} object \jl{f_Ω} and evaluate it at two different \jl{CellPoint} objects \jl{x_Ω} and \jl{x_Γ}. The types \jl{CellField} and \jl{CellPoint} are among the more important ones within the \jl{CellDatum} type-hierarchy.  The \jl{CellPoint} type represents evaluation points defined on the cells of a mesh (including, but not restricted to, integration points). In this case, \jl{x_Ω} represents integration points in the bulk of a Cartesian \ac{fe} mesh, whereas \jl{x_Γ} contains integration points on the boundary faces of the mesh. A \jl{CellPoint} object is composed of three ingredients. First, the plain vector that contains the evaluation points for all cells of the mesh. In particular, \jl{x_Ω} is build from the array \jl{cell_q} that, for each cell, contains a small vector of evaluation points (a single point in this case). The second ingredient is the triangulation object on top of which the integration points are defined. In this case, we build \jl{Ω} (representing the bulk of the mesh) and \jl{Γ} (representing its boundary). The final ingredient of a \jl{CellPoint} is a trait that indicates whether the given coordinates are in the reference or the physical space.\sbc{\footnote{This trait is motivated by the construction of \ac{fe} spaces. \ac{fe} spaces are usually defined in a reference space and then pushed forward to the physical space using the geometrical map. This approach is beneficial in terms of computational cost; many computations can be reused for all cells. However, some spaces must be defined in the physical space.}} Note that we build \jl{x_Ω} from reference coordinates and thus we use \jl{ReferenceDomain()}. The construction of \jl{x_Γ} is analogous. However, the evaluation points in \jl{x_Γ} have 2 components instead of 3 since the reference space for the faces is two-dimensional.

\begin{listing}[ht!]
\juliacode{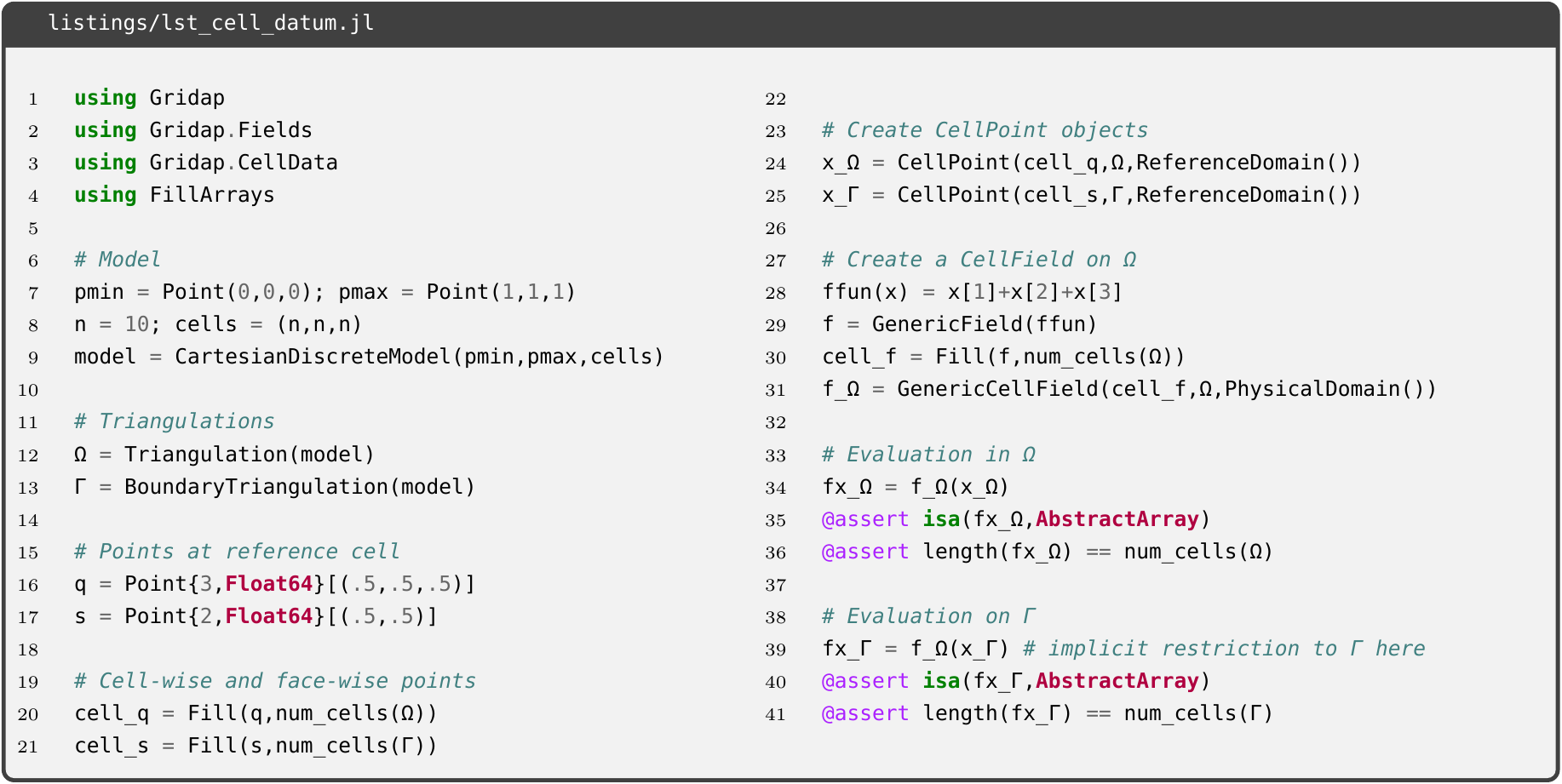}
\caption{Creating and using \jl{CellPoint} and \jl{CellField} objects.}
\label{lst:cell_datum}
\end{listing}

We define the \jl{CellField} object \jl{f_Ω} (which is finally evaluated at \jl{x_Ω} and \jl{x_Γ}) in the remaining part of \lst{lst:cell_datum}. The abstract type \jl{CellField} represents a piece-wise function (or function basis
) defined on the cells of a given mesh.  Here, we show how to manually build a \jl{CellField} from the user-defined function \jl{ffun}. We use the \jl{GenericCellField} constructor that takes the plain array containing a \jl{Field} object for each cell of the mesh. (In this case, all cells share the same field, but one could assign each cell with a different function to have a truly piece-wise function.) The other arguments are the triangulation on top of which the \jl{CellField} is defined and the trait indicating the domain type. In this case, we define the cell field on top of the bulk triangulation \jl{Ω} and we use the value \jl{PhysicalDomain()} since the user-defined function \jl{ffun(x)} is defined in terms of a physical point \jl{x}. The resulting object \jl{f_Ω} can be easily evaluated at the points in \jl{x_Ω} and \jl{x_Γ} using the function notation, namely \jl{f_Ω(x_Ω)} and \jl{f_Ω(x_Γ)}. Safety checks and conversions happen under the hood when evaluating \jl{f_Ω} using the metadata stored in these objects. On the one hand, when calling \jl{f_Ω(x_Ω)}, the code can infer that \jl{f_Ω} is defined in the physical space and the evaluation points \jl{x_Ω} in the reference one. The coordinates in \jl{x_Ω} are internally transformed to the physical space using the geometrical map of \jl{Ω} before their actual evaluation. On the other hand, when calling \jl{f_Ω(x_Γ)}, the code infers that \jl{f_Ω} and \jl{x_Γ} are defined on different triangulations. The restriction of \jl{f_Ω} to the boundary triangulation \jl{Γ} happens implicitly before the evaluation on the points in \jl{x_Γ}. This example clearly shows that it is more convenient to work with \jl{CellDatum} objects than handling the plain low-level arrays directly. Writing the operations that happen under the hood would be much more tedious and less generic. In addition, this high-level \ac{api} does not rely on any code-generation step. It simply leverages the fact that we can efficiently represent cell-wise data using (lazy) arrays.

\subsection{Example: cell-wise elemental matrices (revisited)}

Now, we revisit the example introduced in \lst{lst:cell_mat} and will re-write it using the more flexible syntax provided by \jl{CellDatum} objects (see \lst{lst:cell_mat_bis}). Recall that the objective is to build an array \jl{cell_mat} that represents the elemental stiffness matrix for all cells of a \ac{fe} mesh. In \lst{lst:cell_mat_bis}, we start by building the discrete model using the built-in Cartesian mesh generator, instead of writing the node coordinates and cell connectivity by hand as we previously did in \lst{lst:cell_mat}. The next step is to define the shape functions and the quadrature rule for the reference cell. We build these objects from Gridap functions instead of explicitly doing all the underlying operations. On the one hand, the shape functions are extracted from a \jl{ReferenceFE} object. The abstract type \jl{ReferenceFE} represents a reference \ac{fe} in the sense of Ciarlet \cite{Ciarlet2002}. In particular, the reference \ac{fe} provides the cell topology, a basis of the polynomial space defined on the reference cell (e.g., the shape functions) and the corresponding basis of \acp{dof}. In this example, we build a scalar first-order Lagrangian reference \ac{fe} on top of the reference quadrilateral (represented by the constant \jl{QUAD}) and extract the reference shape functions. Note that the returned object \jl{s} is equivalent to the one computed more manually in \lst{lst:cell_mat}. On the other hand, the quadrature points and weights are computed from a quadrature rule object constructed with the \jl{Quadrature} constructor. Here, we build a quadrature rule of degree 1 for the reference quadrilateral. The resulting arrays \jl{q} and \jl{w} of reference quadrature points and weights are also the same as the ones constructed manually in \lst{lst:cell_mat}.

\begin{listing}[ht!]
\juliacode{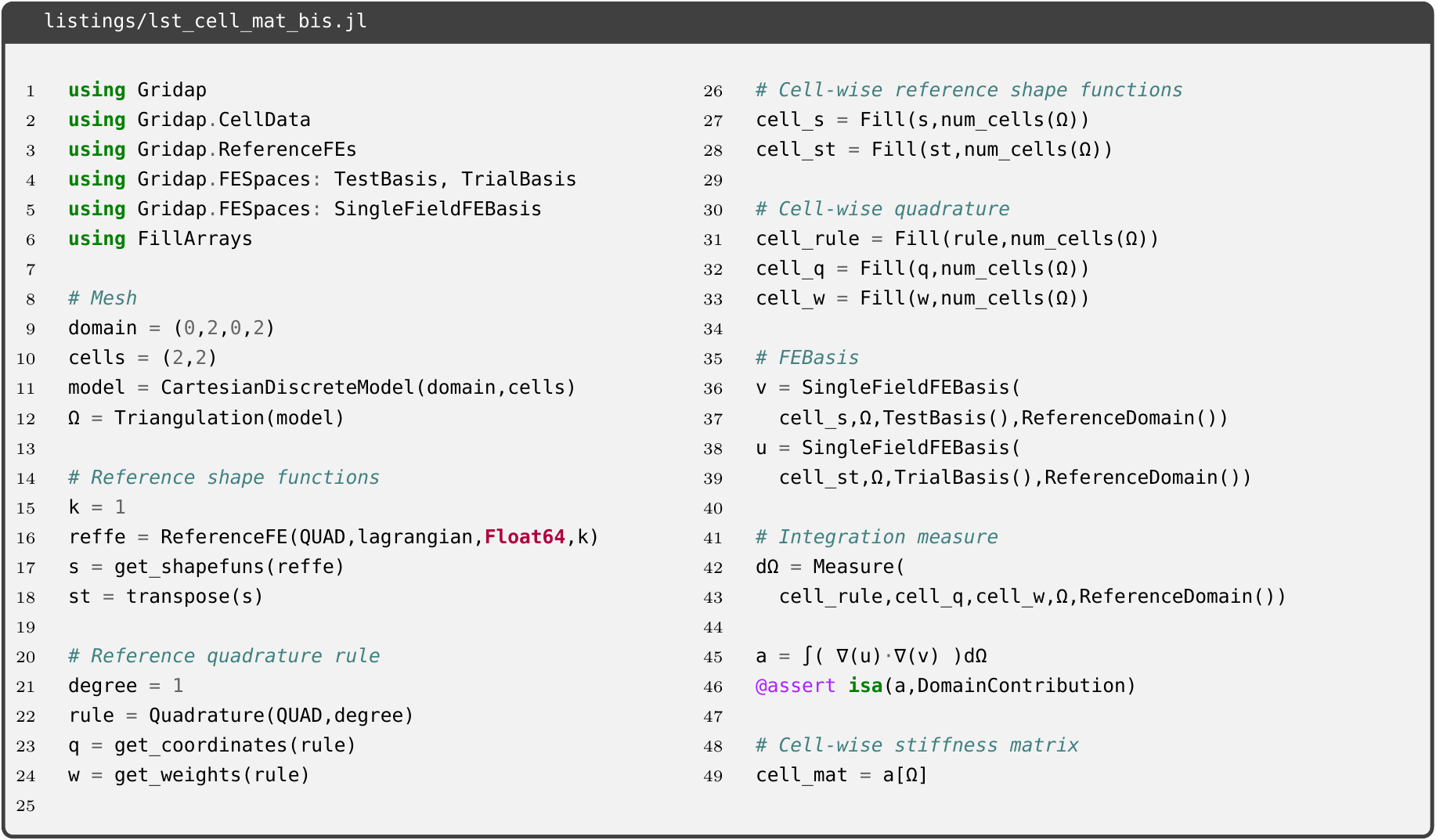}
\caption{Creating a \emph{lazy} vector of elemental stiffness matrices via \jl{CellDatum} objects.}
\label{lst:cell_mat_bis}
\end{listing}

The next part of the code consists in wrapping all these quantities in \jl{CellDatum} objects to compute the elemental matrices. The array of cell-wise shape functions are wrapped in an instance of the concrete type \jl{SingleFieldFEBasis}. This type is a specialization of \jl{CellField} for the particular case of \ac{fe} basis in single-field computations. Even though not exposed in this example, functionality for multi-field problems is also provided by Gridap and will be presented later in \sect{sec:multifield}. The \jl{SingleFieldFEBasis} constructor takes the plain array of cell-wise shape functions, the triangulation object \jl{Ω} on top of which the shape functions are defined, a trait indicating if the shape functions are \emph{test} or \emph{trial} bases and a trait indicating if the shape functions are defined in the reference or the physical space (as commented above). Test shape functions are constructed from \jl{s} (a column vector), whereas trial shape functions are built from the transpose of \jl{s} (a row vector). In this way, we end up having a matrix when broadcasting a product over the row and column vectors as previously shown in \lst{lst:cell_mat}.
 On the other hand, we build a \jl{Measure} object to numerically compute the integrals in the cells of the triangulation \jl{x_Ω}. The \jl{Measure} object is built from the arrays containing the quadrature rule for all cells in the triangulation, the corresponding triangulation object, and the trait indicating that the quadrature rule is in the reference space.

Using the resulting \ac{fe} bases \jl{u} and \jl{v} and the measure object \jl{dΩ}, computing the elemental stiffness matrix is trivial by using the high-level API associated with them. Note that the resulting syntax is almost identical to the corresponding mathematical notation used to define the bilinear form of the Poisson equation, namely $a(u,v)=\int_{\Omega} \nabla u\cdot\nabla v \ {\rm d}\Omega$. Function \jl{∇} returns a new \jl{CellField} object representing the (physical) gradient of the input. In this case, the code detects that the given shape functions \jl{u} and \jl{v} are in the reference space (thanks to the trait \jl{ReferenceDomain()}). The reference shape function gradients are internally pushed to the physical space by using the inverse of the Jacobian of the geometrical map associated with the underlying triangulation. The resulting objects are multiplied using the dot product, which results in a new \jl{CellField} object representing the result of the operation. Finally, we integrate the result using the measure \jl{dΩ}. Note that \jl{∫(f)dΩ} or \jl{∫(f)*dΩ} is just a short-hand for \jl{integrate(f,dΩ)}, which is the function in charge to compute the cell-wise integral of \jl{f} using the measure \jl{dΩ}. The object \jl{a} resulting from the integral is an instance of the type \jl{DomainContribution}. This type is a dictionary-like structure that stores the contributions of several domains to an integral. In this case, the elemental matrices have contributions from a single domain (the triangulation \jl{Ω}), but other examples might have contributions from more than one mesh (e.g., the right-hand-side of a Poisson equation with body forces and Neumann boundary conditions would lead to contributions from two separate domains). To extract the contribution from the cells in the triangulation \jl{Ω}, we simply index \jl{a[Ω]}, which returns a lazy array of length the number of cells in \jl{Ω} containing the contribution to the computed integral for each cell. It is indeed the array containing the elemental stiffness matrices. Even though the code in \lst{lst:cell_mat_bis} is fully functional, it can be further simplified by building the objects \jl{u}, \jl{v}, and \jl{dΩ}  using even higher-level functions instead, see \lst{lst:cell_mat_user}.

\subsection{Finite Element assembly}

We finally show how to assemble global matrices and vectors from the elemental contributions computed in previous examples.  The \ac{fe} assembly routines in Gridap are defined employing the abstract type \jl{Assembler}. This type defines the interface of the assembly routines, which can be specialized for particular cases of interest. Currently, \jl{SparseMatrixAssembler}, which implements an assembler that builds global sparse matrices from elemental contributions, is the only specialization, but others can be added in the future. E.g., one could easily implement a new \jl{Assembler} type that builds \emph{matrix-free} operators \cite{Kiran2020} instead of sparse matrices.

Conceptually, an \jl{Assembler} object takes some elemental matrices or vectors and returns a globally assembled object (e.g., a sparse matrix or a vector). To illustrate this, let us consider the code in \lst{lst:cell_mat_user}. In the first lines, we build the array \jl{cell_mat} containing the elemental stiffness matrices of all the cells in the underlying mesh. This array is equivalent to the one previously computed in \lst{lst:cell_mat_bis}, but in this case, we have used higher-level functions of the Gridap API. In particular, the \ac{fe} basis functions \jl{u} and \jl{v} are extracted from the \jl{FESpace} object \jl{V}. In short, \jl{FESpace} is an abstract type that represents an interpolation space defined on top of the cells of a \ac{fe} mesh. This data structure contains detailed information about the underlying interpolation, e.g., one can recover the test and trial shape functions \jl{v} and \jl{u} from it.  The objective of this snippet is to assembly the local matrices in \jl{cell_mat} into a global sparse matrix \jl{A}. First, we need to define the local-to-global \ac{dof} map. This data is available in the array \jl{cell_dofs}, which can be extracted from the \ac{fe} space \jl{V}. Note that \jl{cell_dofs} is a vector of vectors such that \jl{cell_dofs[e]} returns a vector of global \ac{dof} ids corresponding to cell id \jl{e}. 
The final assembly is straightforward once the arrays \jl{cell_mat} and \jl{cell_dofs} are available. One needs to build an assembler object and then call the \jl{assemble_matrix} function. This function implements the assembly loop and, thus, the user does not need to write it by hand. The assembly of the right-hand-side vector is handled analogously, but it is not included here for the sake of brevity. Note that the assembly interface is flexible enough to potentially use different local-to-global \ac{dof} maps for the rows and columns (useful, e.g., for multi-physics problems). It can also build the assembled operator from several arrays of local contributions (useful, e.g., to assemble operators defined via contributions associated with different integration domains). In addition, the assembly routines are generic and completely independent of the way the local contributions and the local-to-global \ac{dof} maps are constructed. These quantities only need to be instances of sub-types of the Julia \jl{AbstractArray} interface. Thus, one could even assemble local contributions computed with other Julia packages or directly in the user code if needed, which shows the high modularity of the library.

\begin{listing}[ht!]
\juliacode{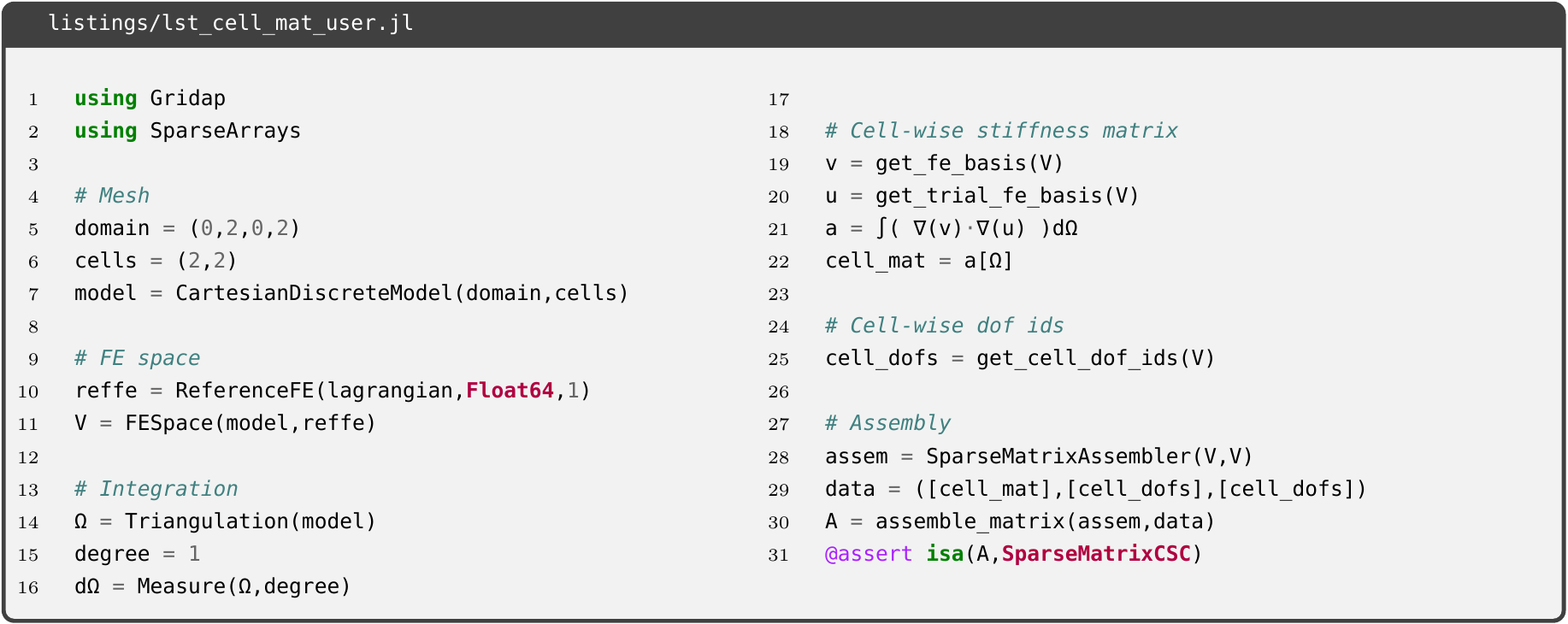}
\caption{Creating a \emph{lazy} vector of elemental stiffness matrices via the high-level API of Gridap and assembling it into a global sparse matrix.}
\label{lst:cell_mat_user}
\end{listing}

\subsection{Putting all pieces together: A Poisson FEM solver explained}
\label{sect:poisson_driver}

At this point, we are ready to combine all pieces discussed in previous sections to build an actual \ac{fe} solver. To illustrate the underlying steps, we consider a Poisson equation as the model problem. This choice is just for simplicity and brevity. Gridap can readily handle much more complex \acp{pde}, including multi-field, nonlinear and time-dependent ones. E.g., we refer to the Gridap tutorial web page \cite{gridap_tutorials_web} for a wide range of examples solved with the library. Here, the goal is to find a numerical approximation of the solution $u$ of a Poisson equation with both Dirichlet and Neumann boundary conditions, namely
\begin{equation}
\left\lbrace
\begin{aligned}
-\Delta u  = f_\Omega  &\text{ in } \Omega,\\
u = f_{\rm D} &\text{ on } \Gamma_{\rm D},\\
n\cdot\nabla u = f_{\rm N} &\text{ on } \Gamma_{\rm N}.
\end{aligned}
\right.
\label{eq:poisson}
\end{equation}
The problem data are the computational domain $\Omega$, the Dirichlet and Neumann boundaries, namely $\Gamma_{\rm D}$ and $\Gamma_{\rm N}$, respectively, and their corresponding boundary values, described by functions $f_\Omega$, $f_{\rm D}$ and $f_{\rm N}$. We define $\Omega\doteq\Omega_\square\setminus\Omega_\circ$ as the Boolean difference of the box $\Omega_\square\doteq(0,.5)\times(0,1)\times(0,1.5)$ and the cylinder $\Omega_\circ\doteq\{x\in\mathbb{R}^3:\ (x_2-.5)^2+(x_3-.75)\leq .3^2\}$. The loading functions $f_\Omega$, $f_{\rm D}$ and $f_{\rm N}$ are defined so that the manufactured function $u(x) = (x_1+x_2+x_3)^k$, $k$ being the interpolation order, solves the problem. This manufactured solution will be used to evaluate the discretization error and to assess the correctness of the solver. 
We consider a standard \ac{fe} method with conforming second-order ($k=2$) Lagrangian elements to compute the numerical solution (see, e.g., \cite{Johnson2009} for more details). The corresponding weak form of the (discrete) problem is: find: $u_h\in U_h$ such that $a(u_h,v_h)=\ell(v_h)$ for all $v_h\in V_h$, where
\begin{equation}
a(u,v) \doteq \int_\Omega \nabla	u\cdot\nabla v \text{ d}\Omega \qquad \text{ and } \qquad \ell(v) \doteq \int_\Omega f_\Omega\	v\text{ d}\Omega + \int_{\Gamma_{\rm N}} f_{\rm N}\ v\text{ d}\Gamma_{\rm N}
\end{equation}
are the forms of the problem, and $U_h$ and $V_h$ are the Lagrangian \ac{fe} spaces that fulfill the Dirichlet conditions $u = f_{\rm D}$ and $u = 0$, respectively. This weak problem can be easily built and solved using the high-level user API of Gridap as shown in \lst{lst:poisson}.

\begin{listing}[ht!]
\juliacode{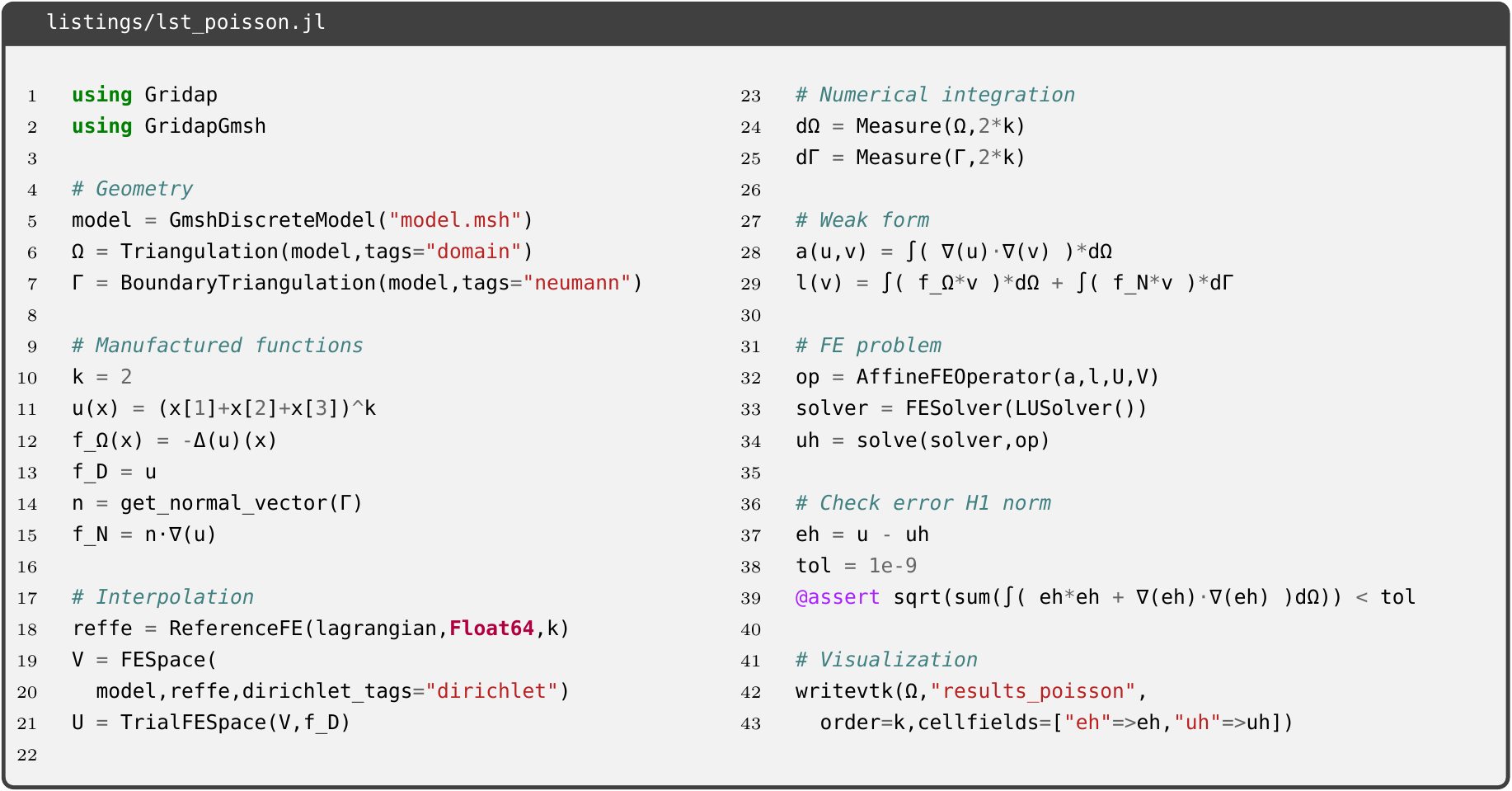}
\caption{Solving a Poisson equation with the Gridap API.}
\label{lst:poisson}
\end{listing}

We start by generating the  \ac{fe} mesh, a.k.a. the discrete model. So far, we have considered only the built-in Cartesian mesh generator for the sake of simplicity. In this example, we show how to deal with more complex domains represented by unstructured grids. The Gridap library outsources the generation of unstructured meshes to well-known mesh generators like Gmsh \cite{Geuzaine2009}. In this example, we consider the plugging GridapGmsh to read a mesh file generated by Gmsh and convert it into a discrete model usable within a Gridap computation. The mesh file is generated using \lst{lst:model} via the official Julia API of the Gmsh project. As a result, it is possible to both generate the mesh (with Gmsh) and solve the \ac{pde} (with Gridap) in the same Julia environment, which is very convenient. Note that we have defined some \emph{physical groups} in the mesh file called \jl{"domain"}, \jl{"dirichlet"}, and \jl{"neumann"}, which identify the domains $\Omega$, $\Gamma_{\rm D}$ and $\Gamma_{\rm N}$ respectively. These labels are used to build the \jl{Triangulation} objects \jl{Ω} and \jl{Γ}, representing $\Omega$ and $\Gamma_{\rm N}$, respectively.

\begin{listing}[ht!]
\juliacode{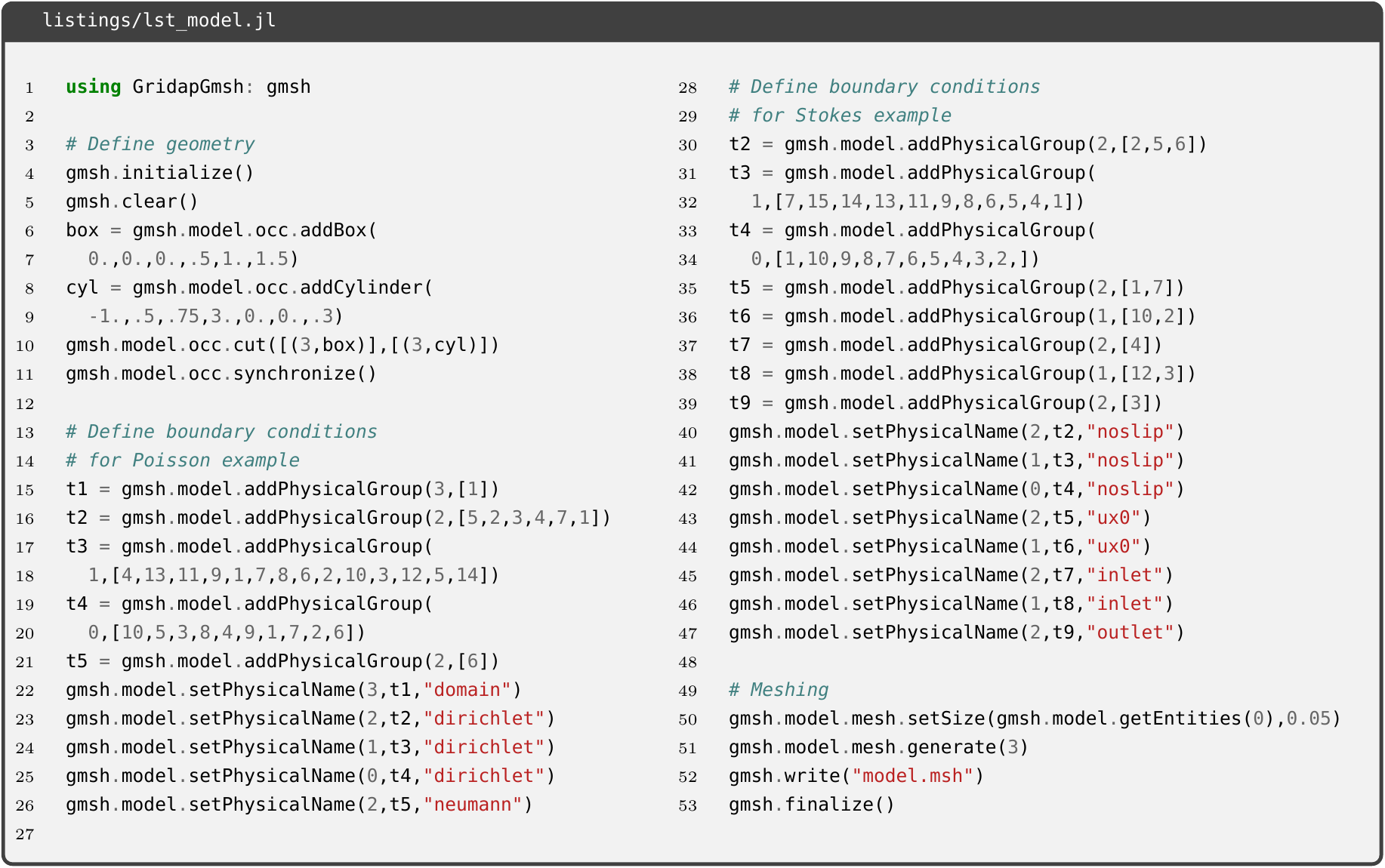}
\caption{Generating the computational mesh used in \lst{lst:poisson} with the Gmsh Julia API.}
\label{lst:model}
\end{listing}

The next step in the \ac{fe} solver is the definition of the manufactured solution $u$ and the corresponding loading functions $f_\Omega$, $f_{\rm D}$ and $f_{\rm N}$. Note that the manufactured solution is defined as a Julia function \jl{u} by using the standard Julia syntax to define function objects. It is not the case for \ac{fe} codes based on symbolic \acp{dsl}, which usually force the user to define functions via strings containing the function definition expressed in C/C++ instead of the programming language of the front-end. However, Gridap allows the user to write arbitrarily complex functions directly in the front-end without compromising performance by leveraging the Julia \ac{jit} compiler, which will be in charge of compiling efficient machine code. From the manufactured solution \jl{u}, once can easily define the loading functions $f_\Omega$, $f_{\rm D}$ and $f_{\rm N}$ thanks to the build-in automatic differentiation capabilities of Gridap. In particular, \jl{Δ(u)} and \jl{∇(u)} compute the Laplacian and the gradient of function \jl{u} via automatic differentiation. The next lines in the code snippet are similar to the ones seen in previous sections. We build the interpolation space \jl{V} and \jl{U} and the measure objects \jl{dΩ} and \jl{dΓ} associated with the triangulations \jl{Ω} and \jl{Γ} respectively. Note that Dirichlet boundary conditions are specified by the name of the physical group \jl{"dirichlet"} describing the Dirichlet boundary in the \jl{FESpace} constructor. The \acp{dof} located on faces included in the selected physical group will be removed from the system (i.e., they are constrained to zero in the resulting test \ac{fe} space \jl{V}).  Spaces with non-homogeneous Dirichlet values, e.g., the trial space \jl{U}, are defined from a test space plus the user-defined function describing the Dirichlet values. Internally, the passed function is interpolated on the faces of the Dirichlet boundary. A vector of Dirichlet \ac{dof} values is created and stored within \jl{U}, which is used later in the \ac{fe} assembly to modify the right-hand-side accordingly.

\begin{figure}[ht!]
\centering
\begin{subfigure}[b]{0.3\textwidth}
\includegraphics[width=\textwidth]{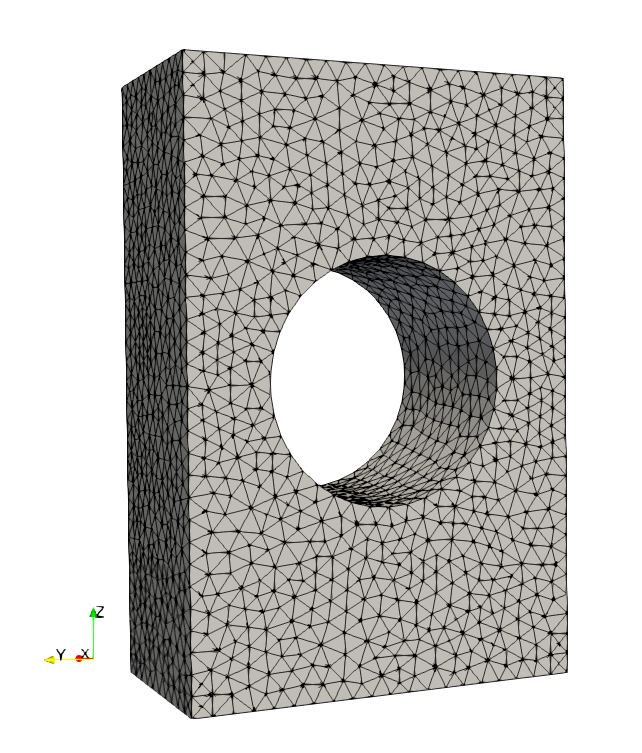}
\caption{Computational mesh \jl{Ω}.}
\end{subfigure}
\begin{subfigure}[b]{0.3\textwidth}
\centering
\begin{scriptsize}
$0.0\cdot 10^0$ \includegraphics[width=0.4\textwidth]{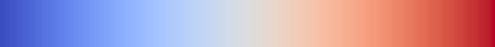} $9.0\cdot 10^0$
\end{scriptsize}

\includegraphics[width=\textwidth]{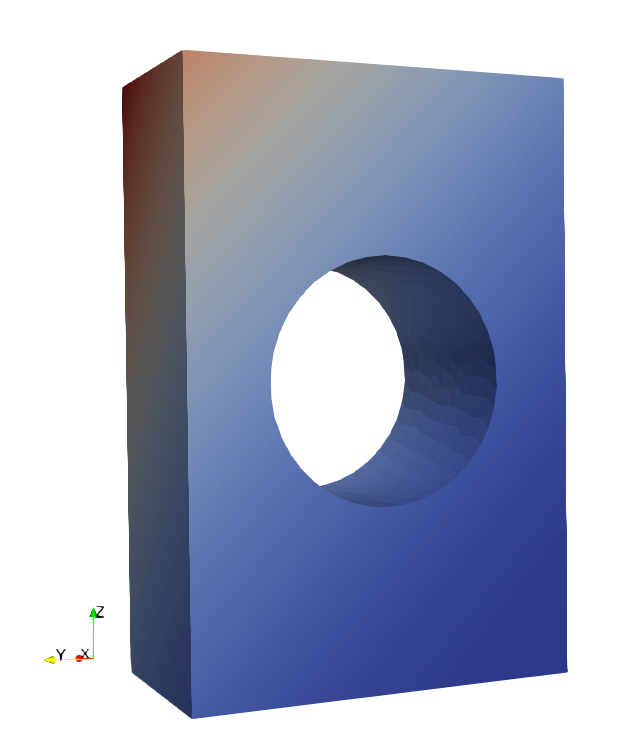}
\caption{Discrete solution \jl{uh} }
\end{subfigure}
\begin{subfigure}[b]{0.3\textwidth}
\centering
\begin{scriptsize}
-$1.8\cdot 10^{-14}$ \includegraphics[width=0.4\textwidth]{fig_colorbar_h.png} $1.5\cdot 10^{-14}$
\end{scriptsize}

\includegraphics[width=\textwidth]{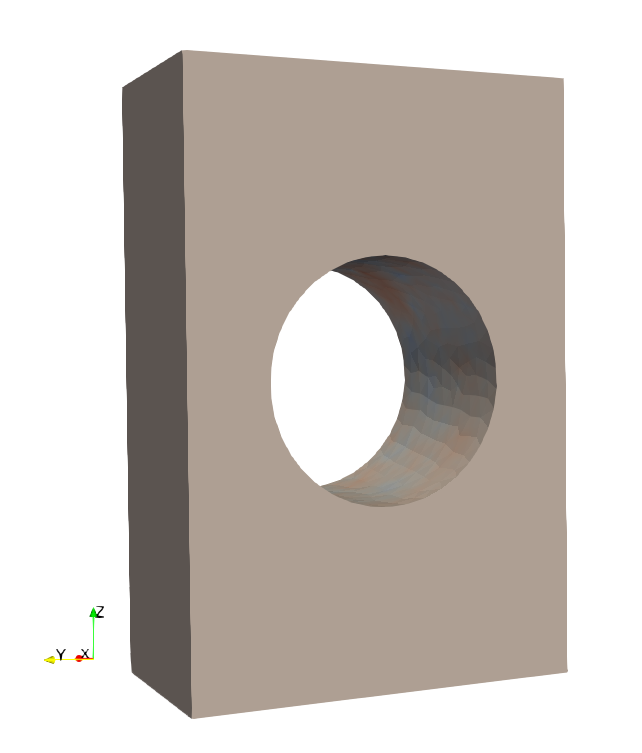}
\caption{Discretization error \jl{eh} }
\end{subfigure}
\caption{Visualization of the simulation implemented in \lst{lst:poisson}.}
\label{fig:driver}
\end{figure}

The resulting cell contributions are assembled into the global matrix and vector, as seen in \lst{lst:cell_mat_user}. The returned object \jl{op} represents the \ac{fe} element problem to be solved, which is a data structure that holds the matrix and vector of the underlying linear system. The \ac{fe} problem is solved with a \jl{FESolver}. It is built from an object implementing the linear solver interface defined in Gridap. This generic interface allows one to use different linear solvers in the same framework. For instance, one can solve the linear system via a $LU$ decomposition from UMFPACK \cite{Davis2004} using \jl{LUSolver}, MKL-Parsido with \jl{PardisoSolver} or PETSc  \cite{petsc-user-ref} with \jl{PetscSolver}. The two latter solver wrappers are available via the plugins GridapParsiso \cite{gridap_pardiso_gh} and GridapPETSc \cite{gridap_petsc_gh},  respectively. The solution of the \ac{fe} problem, namely \jl{uh}, is an instance of the type \jl{FEFunction}, which is a specialization of \jl{CellField} representing a function in a \ac{fe} space. Thus, one can manipulate \jl{uh} via the high-level API implemented for \jl{CellField} objects as we have seen in previous sections. We use this API to compute the interpolation error \jl{eh} and integrate the $H^1$ error norm. The last line of the code snippet writes the discrete solution \jl{uh} and the discretization error \jl{eh} into a VTK file to be visualized in Paraview to produce the plots in \fig{fig:driver}.

%
%
%
%
%
%
%
%
%
%
%
%
%
%
%
%
%
%
%
%
%
%
%

\section{Multi-field equations}
\label{sec:multifield}

\subsection{From single-field to multi-field equations}

In this section, we discuss the main software abstractions that enable the solution of multi-field \acp{pde} with Gridap. This functionality is of high interest since complex applications in science and engineering involve the solution of coupled muli-physics problems and, thus, the handling of single-field equations is insufficient in many cases. We say that a weak form, find $u\in U$ such that $a(u,v)=\ell(v)$ for all $v\in V$, is \emph{multi-field} when the functional spaces $U$ and $V$ are defined as the Cartesian product of \emph{single-field} ones, namely $U=U_1\times\ldots\times U_B$ and $V=V_1\times\ldots\times V_B$. Here, $B$ denotes the number of fields (or blocks) and $U_I$ and $V_I$ for $I=1,\ldots,B$, are the underlying single-field spaces, i.e. functional spaces of scalar, vector, or tensor-valued functions. Note that we will use upper-case letters like $I$ to refer to the blocks/fields and lower-case ones like $i$ to refer to particular entries in an array. Some systems of \acp{pde}, e.g., linear elasticity, are naturally handled by defining vector/tensor-valued single-field spaces. Others, e.g., the Stokes problem, involve several separate fields within a multi-field method. In any case, Gridap is general enough to support both approaches; the former was explained in previous sections, whereas the latter is detailed now.  The Gridap implementation of multi-field \acp{pde} builds on top of the machinery already explained for the single-field case. Thus, it requires the introduction of very few new ingredients. Essentially, we need a new type of \ac{fe} space called \jl{MultiFieldFESpace} that represents the Cartesian product of some single-field spaces and a new container type that preserves the block structure of elemental matrices and vectors in the multi-field case.

\subsection{Recursive block representation of elemental matrices and vectors}

The weak form in multi-field equations has a well-defined block structure. It is associated with the decomposition of  $a$ and $\ell$,
\begin{equation}
a(u,v) = \sum_{I,J=1}^{B} a_{IJ}(u_J,v_I)\text{ and } \ell(v) = \sum_{I=1}^{B} \ell_I(v_I),
\end{equation}
in terms of the underlying single-field quantities. Here $v_I$ represents the component of  $v\in V=V_1\times\ldots\times V_B$ belonging to the single-field space $V_I$ (analogously for $u_J$) and $a_{IJ}$, $\ell_I$ are the individual \emph{blocks} of the weak form. The diagonal terms $a_{II}$ and $\ell_I$ represent the operators and external loads of the underlying single-field equations. The off-diagonal blocks $a_{IJ}$, $I\neq J$,  are the coupling between them. In practice, a significant number of blocks $a_{IJ}$ and $\ell_I$ are potentially equal to zero. (All fields are not necessarily directly coupled, e.g., saddle point systems lead to some zero diagonal blocks.) This \emph{block sparsity} needs to be taken into account to achieve performance. Motivated by this requirement, we adopt the following strategy to represent the elemental contributions to the weak form in multi-field equations.

In Gridap, the elemental matrix and vector $A^e$ and $b^e$ (corresponding to the contribution of a cell id $e$ to the multi-field forms $a$ and $\ell$) are never built explicitly. Instead, we build the elemental matrices and vectors $A^e_{IJ}$ and $b^e_I$ for each of the underlying non-zero single-field forms $a_{IJ}$ and $\ell_I$. By doing so, we reuse the functionality for single-field \acp{pde}, we preserve the block structure of the multi-field problem, we take into account the block sparsity, and we do not need to generate a local \ac{dof} numbering for the multi-field spaces (since the local \ac{dof} numeration of the single-field ones is enough to build the arrays $A^e_{IJ}$ and $b^e_I$). 
%
These arrays are stored in a suitable container type called \jl{ArrayBlock{T,N}}; \jl{T} is the type of the blocks within the object, typically some array type, and \jl{N} is the number of dimensions. For convenience, we define the aliases  \jl{VectorBlock{T}} and \jl{MatrixBlock{T}} for \jl{N=1} and \jl{N=2}, respectively. The \jl{ArrayBlock} type is very simple. It just contains two variables, namely \jl{array} and \jl{touched}. The first one is an array containing the blocks, e.g., the elemental matrices $A^e_{IJ}$ for $I,J=1,\ldots,B$, and the latter is a Boolean mask that indicates which of the entries in \jl{array} are zero and thus have not been computed.

Let us consider \lst{lst:array_block} to illustrate the basic usage of the \jl{ArrayBlock} type. In the first part of the snippet, we build an \jl{ArrayBlock} object that represents three blocks $b_I$, $I=1,2,3$, where $b_2$ and $b_3$ are conceptually zero, and the first block is the array $b_1=(11,12)^\mathrm{t}$. You can interpret this particular case as the local vector in a multi-field \ac{pde} with three fields, where only the external loads of the first field are different from zero. The \jl{ArrayBlock} object is build from the underlying variables \jl{array} and \jl{touched}. On the one hand, \jl{array} is built as a vector of vectors. Note that we only set entry \jl{array[1]}, whereas the other components of \jl{array} remain uninitialized (they are like dangling pointers in C/C++). On the other hand, we define the Boolean mask \jl{touched} indicating that only the first block has been initialized. Note that one does not need to know the sizes of the zero blocks to build an \jl{ArrayBlock} object. This fact enormously simplifies the implementation of multi-field equations in Gridap.  E.g., many contributions of one of the single-fields (e.g. the external loads) can be computed without knowing anything from the other fields, thus allowing to reuse a lot of functionality for single-field problems. Once an \jl{ArrayBlock} is available one can easily inspect its block structure with the \jl{display} function. In this case, the output indicates that the first block is a vector of two entries, whereas the second and third blocks are empty. Several operations are defined over \jl{ArrayBlock} objects. Most of them are optimized to take into account the underlying block sparsity. See for instance the object \jl{c} in \lst{lst:array_block}, which is defined as the difference of two \jl{ArrayBlock} objects. The resulting block sparsity is computed from the given inputs. In this case, the second block is empty since it was empty in both input arrays. A final feature of the \jl{ArrayBlock} implementation worth mentioning is that it is recursive. One can build a block hierarchy by nesting \jl{ArrayBlock} objects, see, e.g., the last lines in \lst{lst:array_block}. This is used in several places in the code, e.g., in \ac{dg} methods for multi-field equations, where one needs two levels of blocks to represent the elemental contributions in interior faces. For instance, for the source term, one needs two blocks $b_L$ and $b_R$ to represent the contributions of the \emph{left} and \emph{right} cell around an interior face, and then one needs yet another block level for each one of the underlying fields.

\begin{listing}[ht!]
\juliacode{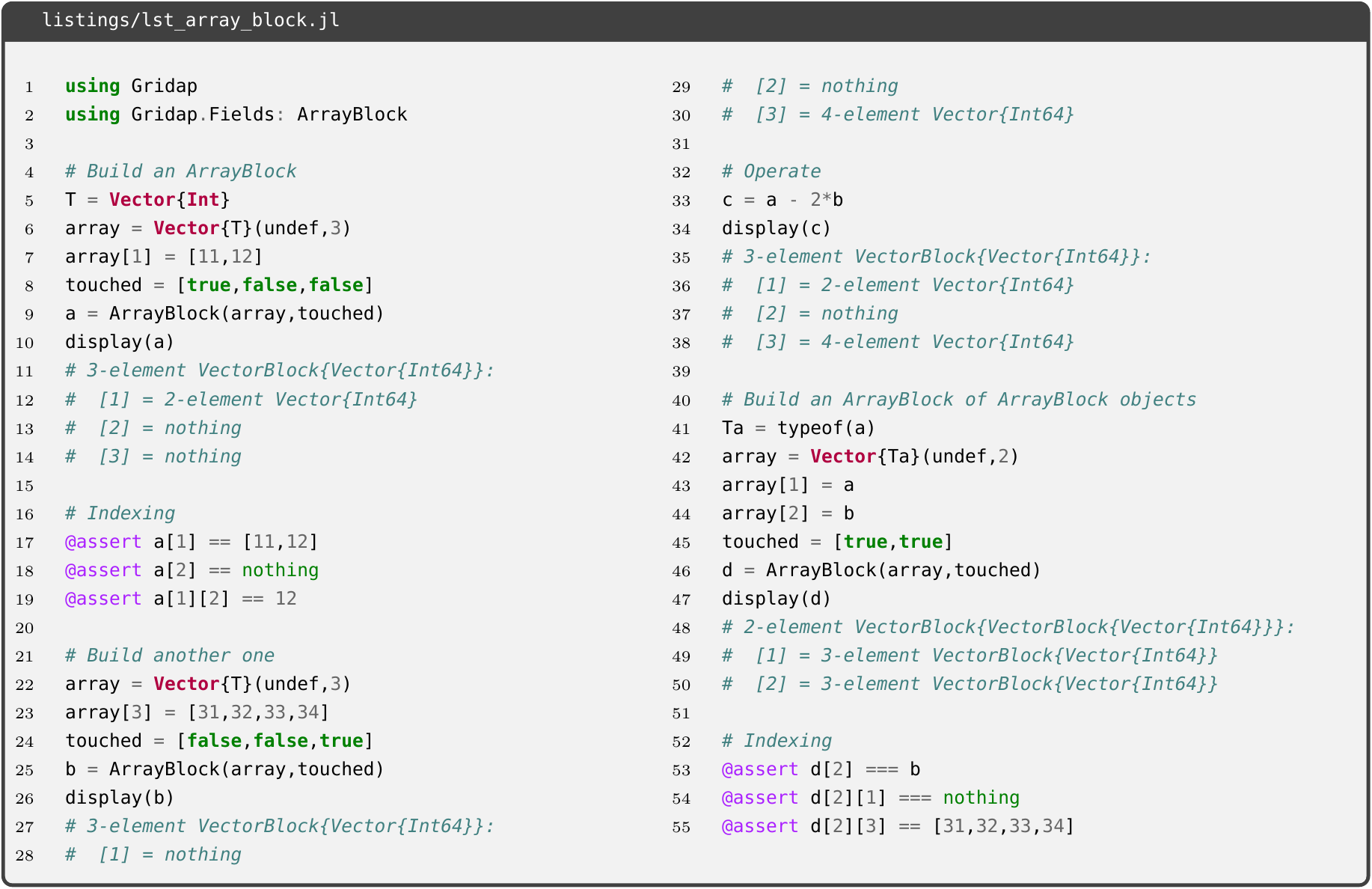}
\caption{Construction and basic usage of \jl{BlockArray} objects.}
\label{lst:array_block}
\end{listing}

%

\subsection{Putting all pieces together: A Stokes FEM solver explained}

Finally, we present the high-level user API to solve multi-field \acp{pde} in Gridap. To this end, we consider a 3D Stokes example as the model problem. The \ac{pde} we want to solve is 
\begin{equation}
\left\lbrace
\begin{aligned}
-\Delta u +\nabla p = f_u  &\text{ in } \Omega,\\
\nabla \cdot u = f_p &\text{ in } \Omega,\\
u = g_{\rm D} &\text{ on } \Gamma_{\rm D},\\
n\cdot\nabla u - pn = g_{\rm N} &\text{ on } \Gamma_{\rm N},\\
\end{aligned}
\right.
\label{eq:stokes}
\end{equation}
where $u$ and $p$ are the velocity and pressure fields, $f_u$ and $f_p$ are the source terms ($f_p=0$ usually), and $g_{\rm D}$ and $g_{\rm N}$ are the Dirichlet and Neumann boundary conditions on the boundaries $\Gamma_{\rm D}$ and $\Gamma_{\rm N}$ respectively.
In this example, the computational domain $\Omega$ is the same as the one considered in the Poisson example in \sect{sect:poisson_driver}, but we define different boundaries to impose boundary conditions as detailed in \tabl{tabl:stokes_bcs}. In particular, we consider a homogeneous Neumann condition on $\Gamma_{\rm N}$ and impose an inflow velocity on $\Gamma_{\mathrm{D}1}$ with value 
\begin{equation}
g_{\mathrm{D}1}(x)\doteq(0,0,(1-(4x_1-1)^2)(1-(2x_2-1)^2))^\mathrm{t}.
\end{equation}
On the other hand, we define a no-slip boundary on $\Gamma_{\mathrm{D}2}$ and constrain the first component of the velocity on $\Gamma_{\mathrm{D}3}$. \tabl{tabl:stokes_bcs} contains also the labels defined in the mesh file in \lst{lst:model}, which represent each of these boundaries. 

\begin{table}[ht!]
\begin{tabular}{llll}
\toprule
Symbol & Label & Definition & Condition\\
\midrule
$\Gamma_{\rm N}$ & \jl{"outlet"} & $(0,0.5)\times(0,1)\times\{1.5\}$ & $n\cdot\nabla u - pn=0$ \\
$\Gamma_{\mathrm{D}1}$ & \jl{"inlet"} & $(0,0.5)\times(0,1)\times\{0\}$ &$u=g_{\mathrm{D}1}$\\
$\Gamma_{\mathrm{D}2}$ & \jl{"noslip"} & $(0,0.5)\times\{0,1\}\times(0,1.5)\cup\partial\Omega_\circ$ & $u=0$\\
$\Gamma_{\mathrm{D}3}$ & \jl{"ux0"} & $\{0,0.5\}\times(0,1)\times(0,1.5)$ &$u_1=0$\\
\bottomrule
\end{tabular}
\caption{Boundary conditions for the Stokes example in \lst{lst:stokes_driver}.}
\label{tabl:stokes_bcs}
\end{table}

The numerical approximation is based on an inf-sup stable Taylor-Hood pair \cite{Brezzi1991} for velocity and pressure fields, with continuous $P_2$ and $P_1$ elements, respectively. For this formulation, stabilization is not needed and the discrete weak form for this problem is: find $(u_h,p_h)\in U_h\times Q_h$ such that $a((u_h,p_h),(v_h,q_h))=l((v_h,q_h))$ for all $(v_h,q_h)\in V_h\times Q_h$, where
\begin{equation}
a((u,p),(v,q)) \doteq \int_\Omega \nabla	u\cdot\nabla v - p(\nabla\cdot v) +  (\nabla\cdot u) q   \text{ d}\Omega \text{ and } \ell(v) \doteq \int_\Omega f_u\	v + f_p\ q\text{ d}\Omega.
\end{equation}
The spaces $U_h$ and $V_h$ are the continuous Lagrangian $P_2$ spaces for the velocity fulfilling the Dirichlet conditions, whereas $Q_h$ is the continuous Lagrangian $P_1$ space for the pressure.

\begin{listing}[ht!]
\juliacode{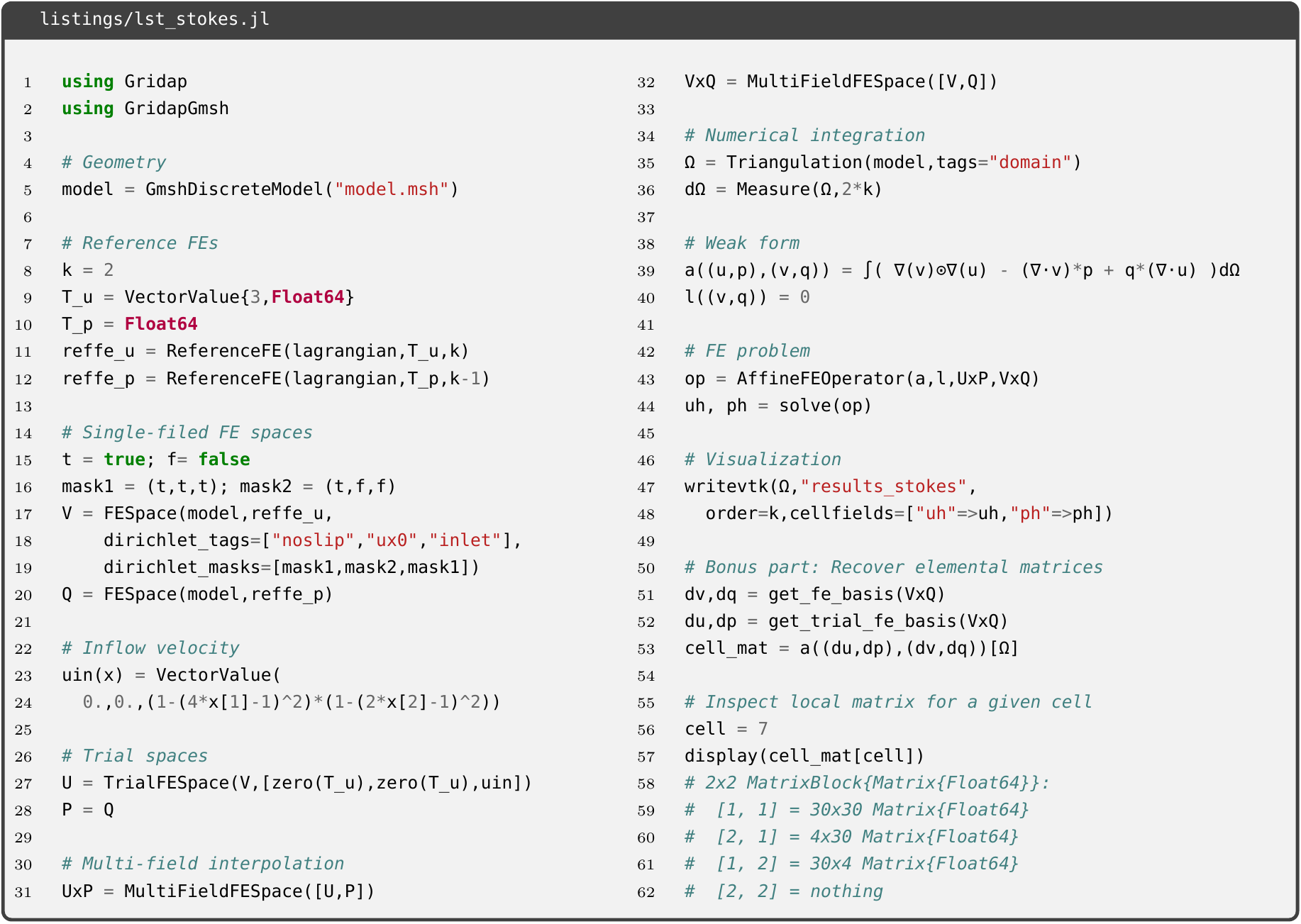}
\caption{Solving a Stokes equation with the Gridap API.}
\label{lst:stokes_driver}
\end{listing}

\begin{figure}[ht!]
\centering
\begin{subfigure}[b]{0.3\textwidth}
\includegraphics[width=\textwidth]{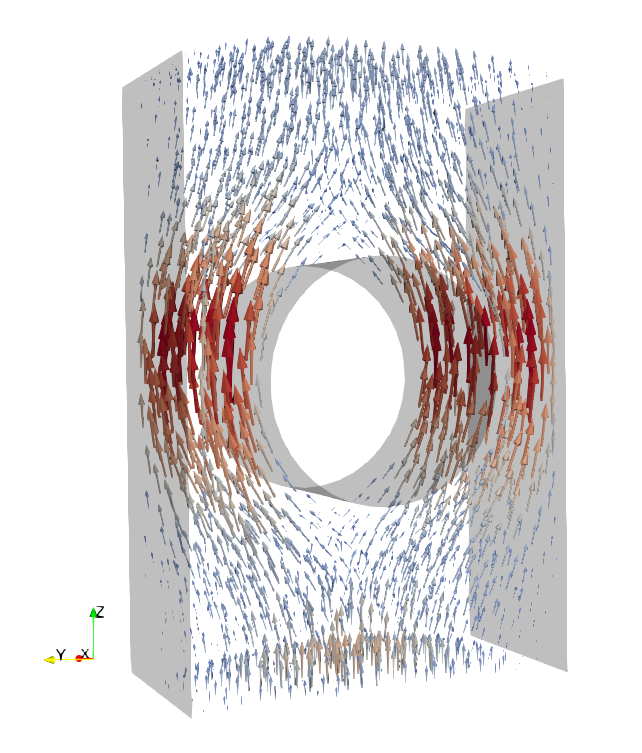}
\caption{}
\end{subfigure}
\begin{subfigure}[b]{0.3\textwidth}
\centering
\begin{scriptsize}
$0.0\cdot 10^0$ \includegraphics[width=0.4\textwidth]{fig_colorbar_h.png} $1.6\cdot 10^0$
\end{scriptsize}

\includegraphics[width=\textwidth]{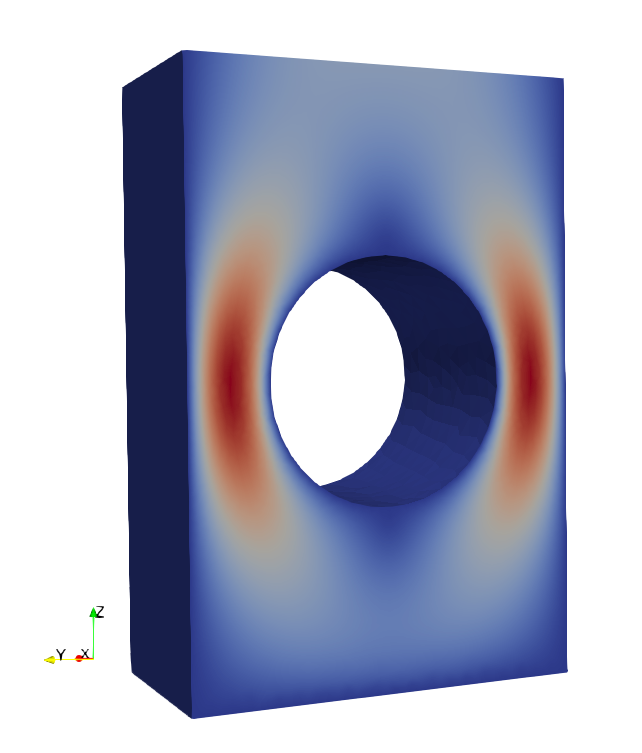}
\caption{}
\end{subfigure}
\begin{subfigure}[b]{0.3\textwidth}
\centering
\begin{scriptsize}
$-4.7\cdot 10^0$ \includegraphics[width=0.4\textwidth]{fig_colorbar_h.png}  $1.7\cdot 10^2$
\end{scriptsize}

\includegraphics[width=\textwidth]{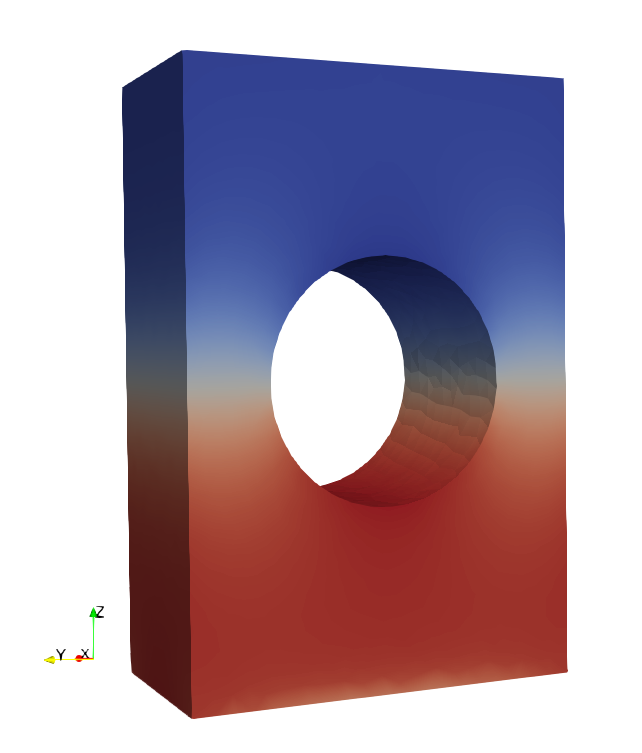}
\caption{}
\end{subfigure}
\caption{Visualization of the Stokes simulation implemented in \lst{lst:stokes_driver}. (a) Velocity field displayed as arrows scaled and colored with velocity magnitude. (b) Velocity field magnitude. (c) Pressure field.}
\label{fig:driver_stokes}
\end{figure}

As for the Poisson example, the high-level API of Gridap allows one to build and solve this problem in a compact way using syntax that resembles the mathematical notation (see \lst{lst:stokes_driver}). The first part of the driver builds the \ac{fe} spaces for velocity and pressure. These spaces are defined using the \ac{api} for single-field problems. The definition of the velocity space requires some extra work since we have three different parts of the Dirichlet boundary where we define different conditions, namely $\Gamma_{\mathrm{D}1}$, $\Gamma_{\mathrm{D}2}$, and $\Gamma_{\mathrm{D}3}$. Note that one can select which components of a vector/tensor-valued space to constrain on the Dirichlet boundary by supplying a vector of masks. Once the single-field spaces are defined, we define the tensor product via the \jl{MultiFieldFESpace} constructor. Using the tensor product spaces, we can proceed with the driver as in the single-field case. The major exception is that one needs to unpack the velocity and pressure components in the arguments of the weak form, e.g., \jl{a((u,p),(v,q))}, leading to a syntax almost identical to the corresponding mathematical notation for the definition of the multi-field bilinear and linear forms. The second difference is that one needs to unpack the velocity and pressure solutions after solving the problem. The resulting objects  \jl{uh} and \jl{ph} are single-field \ac{fe} functions that can be handled as discussed in previous sections. In particular, one can export them in VTK format for visualization in Paraview, see \fig{fig:driver_stokes}.

The last part of \lst{lst:stokes_driver} is not needed to solve the problem but is included here to illustrate how one can inspect the lazy vector containing the elemental matrices. The approach is analogous to the one previously explained for the Poisson equation. One extracts the trial and test \ac{fe} bases and evaluates the bilinear form with them. The vector \jl{cell_mat} contains the local Stokes operator for all cells in the mesh. We inspect the contents of this vector for a particular cell (cell id $7$ in this case). As the output of the \jl{display} function reveals, the local operator is represented with a \jl{MatrixBlock} object with a $2 \times 2$ block structure, as expected. Note that the last block is empty. The weak form has no term in the diagonal pressure block (we have used an inf-sup stable formulation). The code automatically infers the block sparsity from the multi-field weak form, as expected.

\section{Numerical Examples}
\label{sec:examples}

\subsection{Objectives}

The goal of the numerical examples below is to evaluate the performance of the \ac{fe} framework introduced in this paper and available in the Gridap package. The main goal is to demonstrate that the Julia implementation in Gridap can be as efficient as other state-of-the-art \ac{fe} libraries that combine a C/C++ back-end plus with a similar Python front-end. Note that the goal is not to outperform previous \ac{fe} packages but to show that Gridap can achieve similar performance while using a simpler design based on a user-friendly programming language (Julia) without considering sophisticated compilers of variational forms like FFC \cite{Kirby2006}. To this end, we perform performance comparisons against a recent version of FEniCS \cite{Logg2010}, arguably the most popular state-of-the-art \ac{fe} code, which provides a compact high-level user interface similar in spirit to the one in Gridap.

\subsection{Experimental setup} The numerical experiments below are computed on a personal computer with an Intel Core i7-10510U CPU with 1.80GHz and approximately 16GiB of RAM running with Ubuntu 18.04.3 LTS. We use Gridap version 0.16.2 and Julia version 1.6.2. The Julia binary is invoked with flags \jl{--check-bounds=no} and \jl{-O 3} to disable array bound checks and select the maximum optimization level of the compiler, respectively. We use FEniCS version 2019.2.0-dev0 (the latest version available when running the examples) with Python 3.6.9. In addition, we force optimization of the form compiler by setting the parameters as shown in \lst{lst:fenics_ops}. All reported times are computed as the minimum of 4 identical runs to mitigate the effect of operating system jitter. Regarding mesh generation, we consider both structured and unstructured grids.  The structured meshes are generated with the build-in mesh generators available in Gridap and FEniCS, whereas the unstructured meshes are generated with Gmsh version 4.4.1. The \jl{.msh} file generated by Gmsh is directly read with the GridapGmsh plugin (as already shown in \lst{lst:poisson}). In \jl{FEniCS}, the \jl{.msh} file is converted to \jl{xdmf} format with the meshio python package version 4.3.11 \cite{Schlomer2021} before being consumed by FEniCS routines.

\begin{listing}[ht!]
\centering
\includegraphics[width=\textwidth]{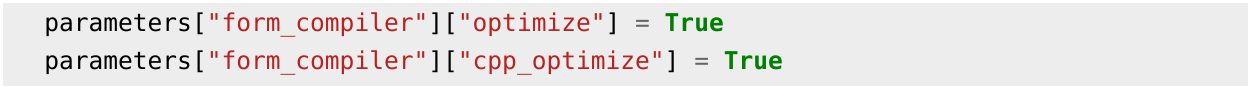}
\vspace*{-2em}
\caption{Form compiler options used in FEniCS computations.}
\label{lst:fenics_ops}
\end{listing}

%
%

\subsection{Poisson benchmark} We consider the Poisson equation as a model problem to assess the performance of Gridap in single-field computations. In this benchmark, the problem formulation is the same as the one in \equ{eq:poisson}, except that we only consider Dirichlet boundary conditions, namely $\Gamma_{\rm D}=\partial\Omega$ and $\Gamma_{\rm N}=\emptyset$. The computational domain $\Omega$ is defined now by two different geometries. The first one is the unit cube $(0,1)^3$ and the second one is the perforated box geometry meshed with \lst{lst:model}. We label these geometries as \emph{geo1} and \emph{geo2}, respectively. The first geometry can be meshed with structured grids. The second case requires unstructured grids. Note that we use both structured and unstructured grids to explore potential performance differences between these cases. The \ac{fe} formulation used to discretize the problem is a continuous Galerkin method with Lagrangian elements of first ($k=1$) and second ($k=2$) order. We only use tetrahedral meshes for the comparison because FEniCS does not support hexahedral meshes. (Note that Gridap supports general cell topologies.) A summary of the parameters used in this example is found in \tabl{tabl:poisson}. The Gridap code used in this benchmark is an adaptation of the one in \lst{lst:poisson}, in which we have included timing directives. The FEniCS code builds on the code examples in the FEniCS project web page. 

\begin{table}[ht!]
\centering
\begin{tabular}{ll}
\toprule
\bf Parameter & \bf Value\\
\midrule
Model problem & Poisson equation with Dirichlet boundary conditions\\
Geometry & Unit cube (geo1) and perforated box (geo2)  \\
Numerical scheme & Continuous Galerkin with Lagrangian interpolation\\
$k$ (interpolation order)  & $1$ and $2$ \\
Cell topology & Tetrahedron\\
Mesh type & Structured for geo1 and unstructured for geo2\\
\ac{fe} library  & Gridap and FEniCS \\
\bottomrule
\end{tabular}
\vspace{1em}
\caption{Summary of the Poisson benchmark.}
\label{tabl:poisson}
\end{table}

We define two different time measures for the performance comparison. The first one is the total time needed to assemble the linear system from scratch. The second time measures the time to re-assemble the linear system by reusing as much data as possible from the previous one. Note that the first timing provides information about the global performance of the library, including all simulation phases needed to build the discrete system of linear equations. It includes reading (or generating) the \ac{fe} mesh and the interpolation spaces, handling the weak form, allocating the system, and assembling it. This timing provides information about the overall Gridap performance and includes several parts of the library not explained in detail herein. The second timing measures the cost of the assembly loop. One can relate this measure to the time needed to assemble the linearized problem at each Newton-Raphson iteration in the solution of a nonlinear \ac{pde} since it is possible to reuse a significant amount of information from previous iterations like the sparsity pattern of the system matrix. This second measure is directly related to the low-level routines detailed in this paper. These two time measures are labeled for further reference as \emph{assembly from scratch} and \emph{in-place assembly} respectively.

\begin{figure}[ht!]
\centering

\begin{small}
\includegraphics[width=0.04\textwidth]{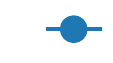} $k=1$ Gridap
\includegraphics[width=0.04\textwidth]{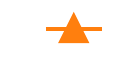} $k=2$ Gridap
\includegraphics[width=0.04\textwidth]{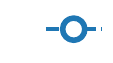} $k=1$ FEniCS
\includegraphics[width=0.04\textwidth]{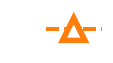} $k=2$ FEniCS
\end{small}

\vspace{0.5em}

\begin{subfigure}{0.49\textwidth}
\centering
\includegraphics[width=0.49\textwidth]{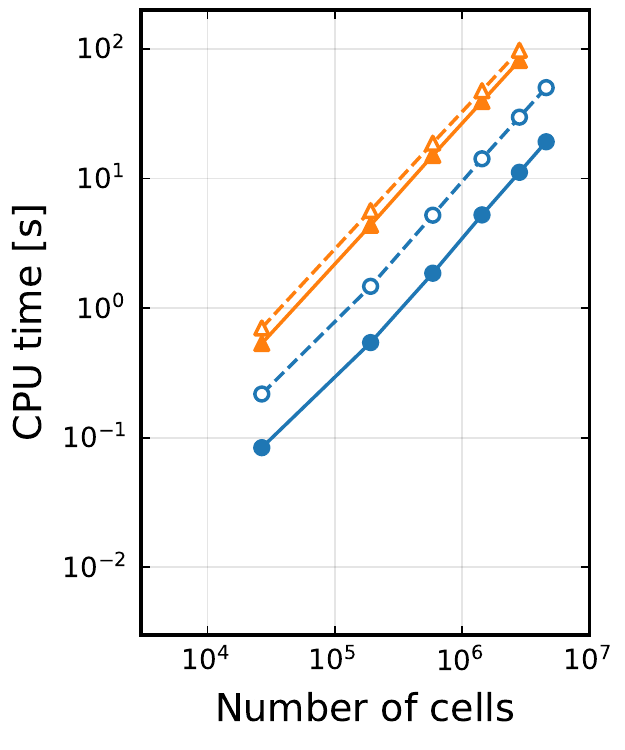}
\includegraphics[width=0.49\textwidth]{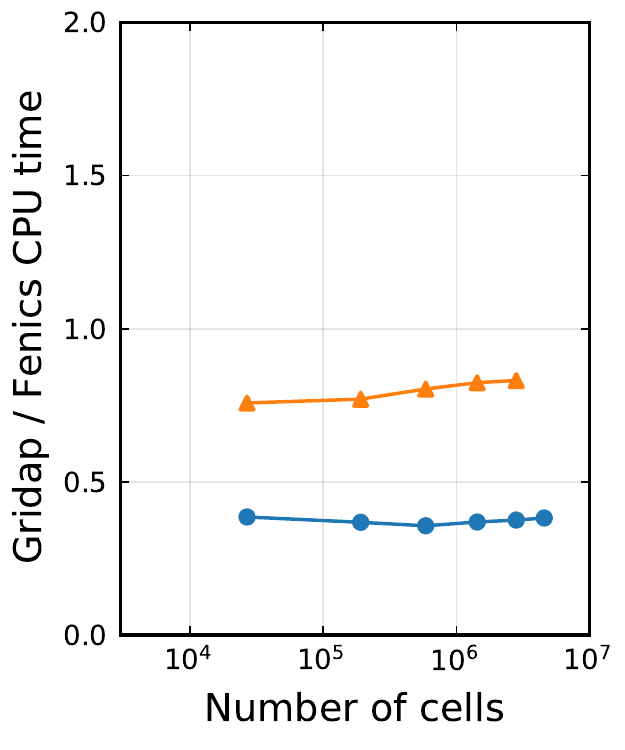}
\caption{Assembly from scratch with unstructured mesh.}
\end{subfigure}
\begin{subfigure}{0.49\textwidth}
\centering
\includegraphics[width=0.49\textwidth]{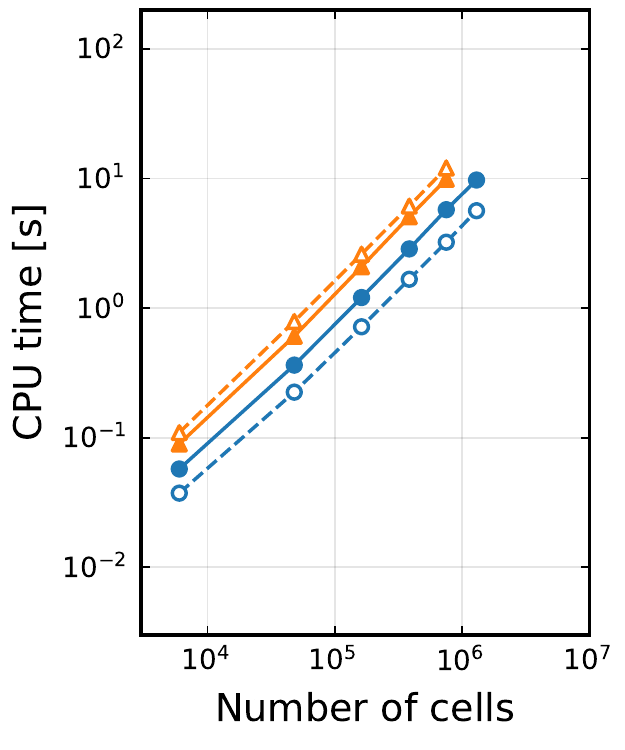}
\includegraphics[width=0.49\textwidth]{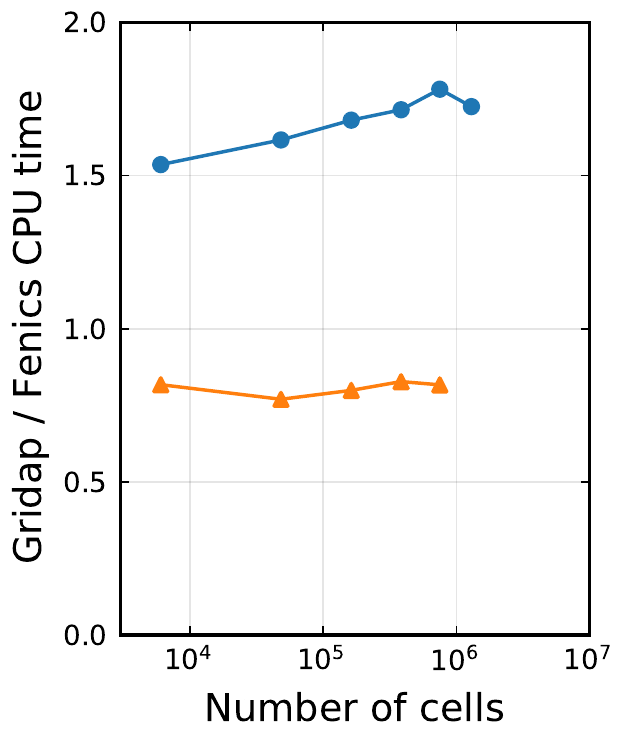}
\caption{Assembly from scratch with structured mesh.}
\end{subfigure}

\begin{subfigure}{0.49\textwidth}
\centering
\includegraphics[width=0.49\textwidth]{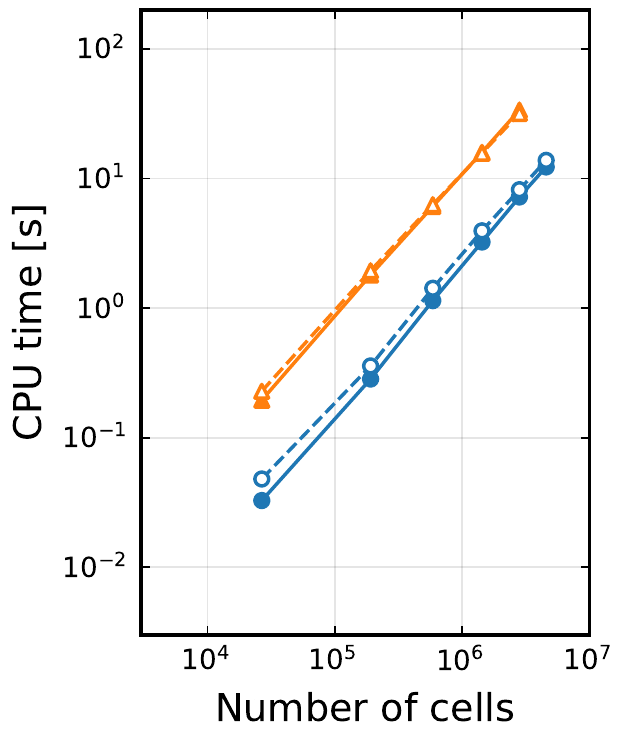}
\includegraphics[width=0.49\textwidth]{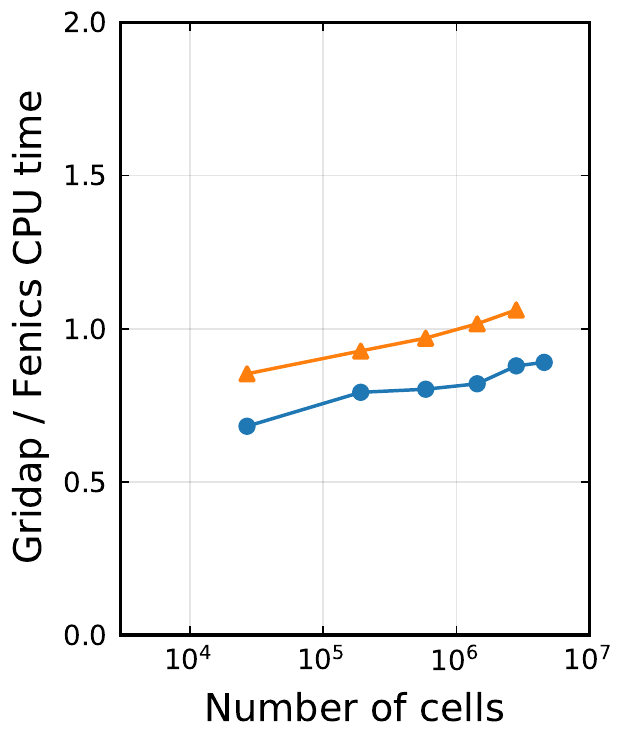}
\caption{In-place assembly with unstructured mesh.}
\end{subfigure}
\begin{subfigure}{0.49\textwidth}
\centering
\includegraphics[width=0.49\textwidth]{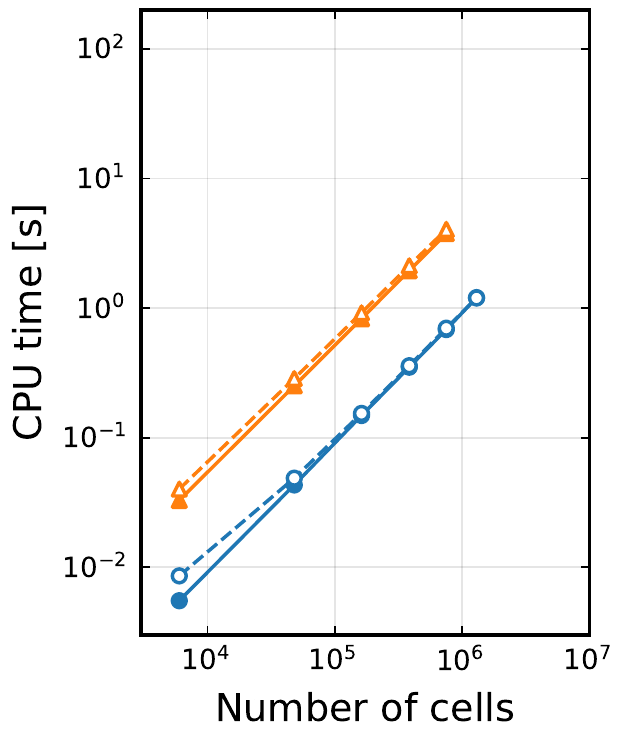}
\includegraphics[width=0.49\textwidth]{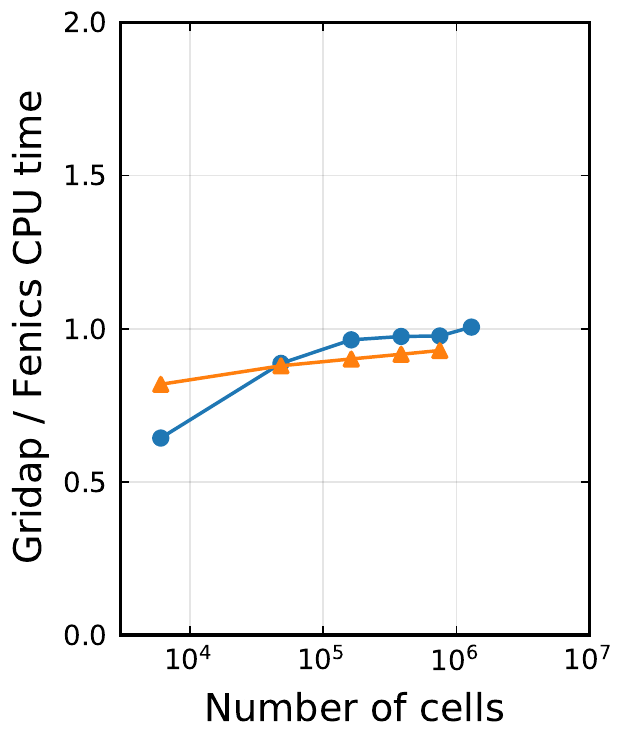}
\caption{In-place assembly with structured mesh.}
\end{subfigure}
\caption{Poisson benchmark: Scaling of CPU time for the phases \emph{assembly from scratch} and \emph{in-place assembly} for computations on unstructured mesh and structured meshes.}
\label{fig:poisson_cmp_times}
\end{figure}

\fig{fig:poisson_cmp_times} shows the scaling of these two timing phases with respect to the number of cells in the mesh for both structured and unstructured meshes. We also provide Gridap times divided by  FEniCS ones to facilitate the comparison between the two libraries. Gridap is slightly faster in most cases but differences are not substantial. The main differences are in the assembly from scratch phase. For $k=1$, Gridap is faster for unstructured grids, whereas FEniCS is faster for structured ones. Note that Gridap is optimized for Cartesian grids of n-cube cells but not for structured grids of simplices. This latter case uses generic data structures for unstructured grids, which are not leveraging any particular grid topology. Optimization for this specific case would reduce the difference between the two libraries. In addition, FEniCS is taylored for simlicial elements, whereas Gridap implements general cell topologies. In any case, the more generic implementation of Gridap does not result in a significant performance hit with  respect to FEniCS.

\subsection{Stokes benchmark}

In this second benchmark, we consider the Stokes equation to assess the performance of Gridap in multi-field computations. We consider \equ{eq:stokes} with pure Dirichlet boundary conditions, i.e., $\Gamma_{\rm} = \partial\Omega$ and $\Gamma_{\rm N}=\emptyset$, and define the forcing functions $f_u$, $f_p$, and $g_{\rm D}$ so that the manufactured velocity and pressure functions $u(x) = (x^2_1+2x^2_2,-x^2_2,0)^{\rm t}$ and $p(x) = x_1 + 3x_2$ are the solutions of the problem.
Since we impose Dirichlet boundary conditions on the entire boundary $\partial\Omega$, the mean value of the pressure is constrained to zero to have a well-posed problem,\footnote{{We note that Gridap provides a dedicated finite element space that internally handles the zero mean constraint. It is useful when defining pressure spaces in confined flow problems (i.e., when the full boundary is of Dirichlet type like in this benchmark).}}
\begin{equation}
\int_\Omega p \text{ d}\Omega = 0.
\label{eq:zeromean}
\end{equation}
We define the computational domain $\Omega$ with the same geometries (the unit cube and the perforated box labeled as \emph{geo1} and \emph{geo2} as before) and the same \ac{fe} meshes (tetrahedral cells) as previously considered in the Poisson benchmark. The \ac{fe} method used to discretize the problem is the one considered in the Stokes example in \lst{lst:stokes_driver}. The definition of this benchmark is summarized in \tabl{tabl:stokes}.

\begin{table}[ht!]
\centering
\begin{tabular}{ll}
\toprule
\bf Parameter & \bf Value\\
\midrule
Model problem & Stokes equation with Dirichlet boundary conditions\\
Geometry & Unit cube (geo1) and perforated box (geo2)  \\
Numerical scheme & Taylor-Hood interpolation\\
$k$ (interpolation order)  & $2$ for velocity and $1$ for pressure \\
Cell topology & Tetrahedron\\
Mesh type & Structured for geo1 and unstructured for geo2\\
\ac{fe} library  & Gridap and FEniCS \\
\bottomrule
\end{tabular}
\vspace{1em}
\caption{Summary of the Stokes benchmark.}
\label{tabl:stokes}
\end{table}

%

\begin{figure}[ht!]
\centering

\begin{small}
\includegraphics[width=0.04\textwidth]{fig_legend_gridap_2.pdf} Gridap
\includegraphics[width=0.04\textwidth]{fig_legend_fenics_2.pdf}  FEniCS
\end{small}

\vspace{0.5em}

\begin{subfigure}{0.49\textwidth}
\centering
\includegraphics[width=0.49\textwidth]{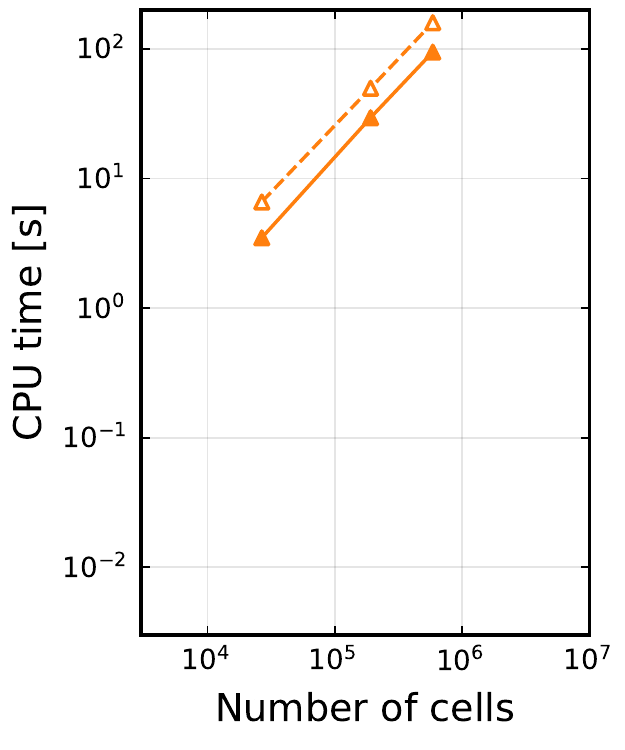}
\includegraphics[width=0.49\textwidth]{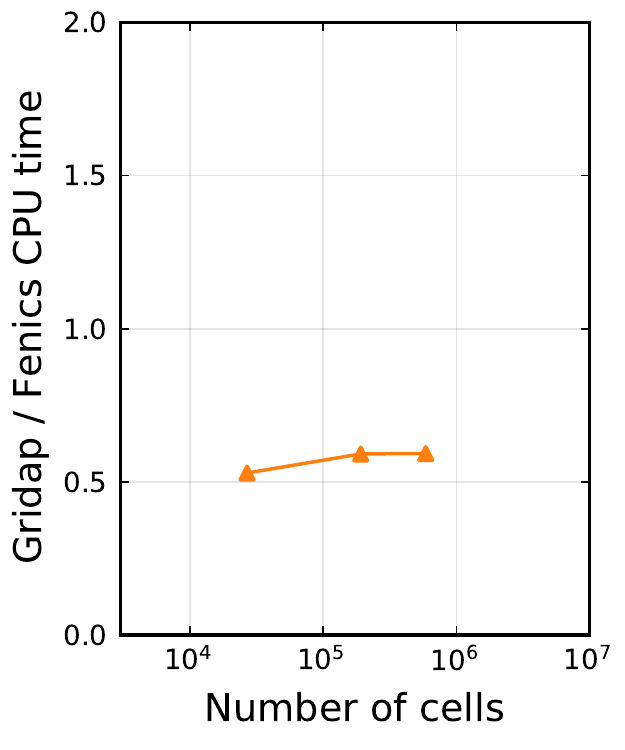}
\caption{Assembly from scratch with unstructured mesh.}
\end{subfigure}
\begin{subfigure}{0.49\textwidth}
\centering
\includegraphics[width=0.49\textwidth]{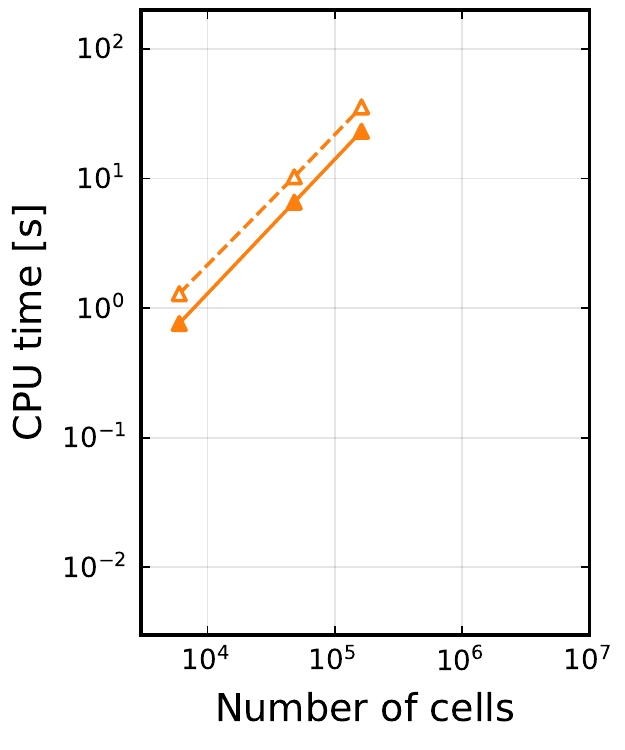}
\includegraphics[width=0.49\textwidth]{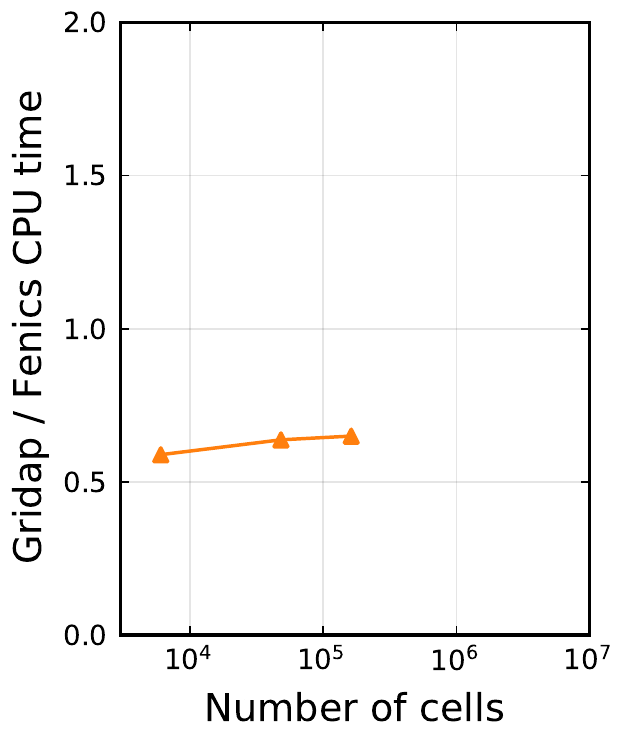}
\caption{Assembly from scratch with structured mesh.}
\end{subfigure}

\begin{subfigure}{0.49\textwidth}
\centering
\includegraphics[width=0.49\textwidth]{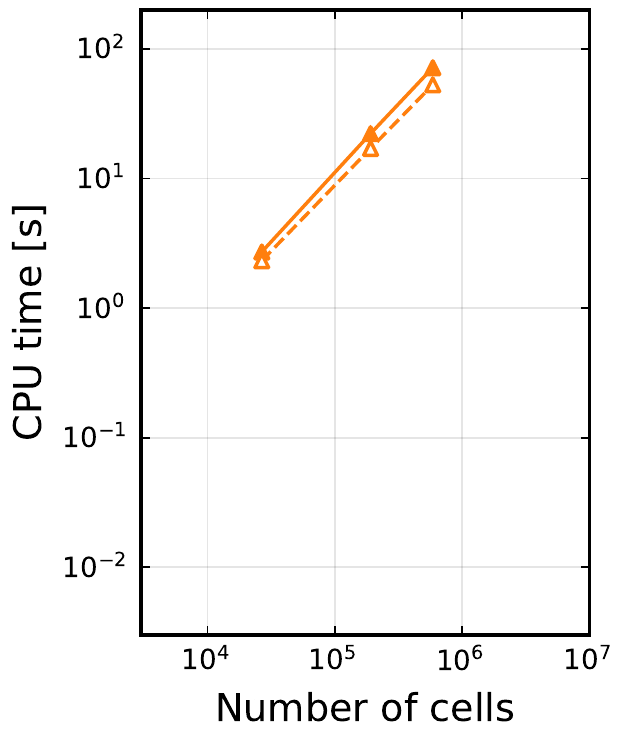}
\includegraphics[width=0.49\textwidth]{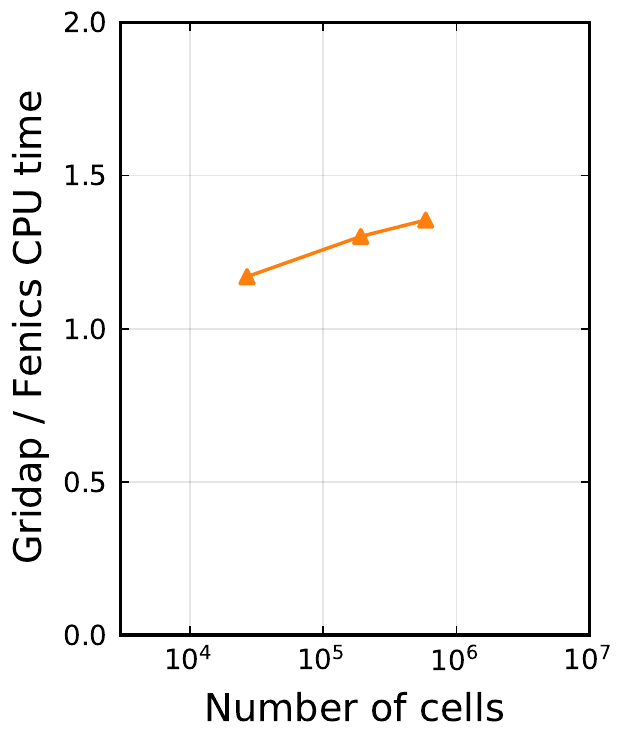}
\caption{In-place assembly with unstructured mesh.}
\end{subfigure}
\begin{subfigure}{0.49\textwidth}
\centering
\includegraphics[width=0.49\textwidth]{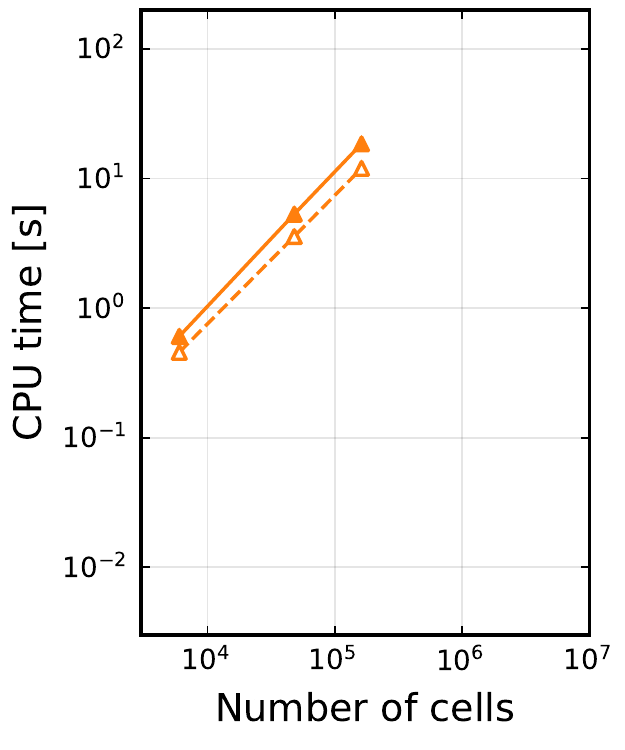}
\includegraphics[width=0.49\textwidth]{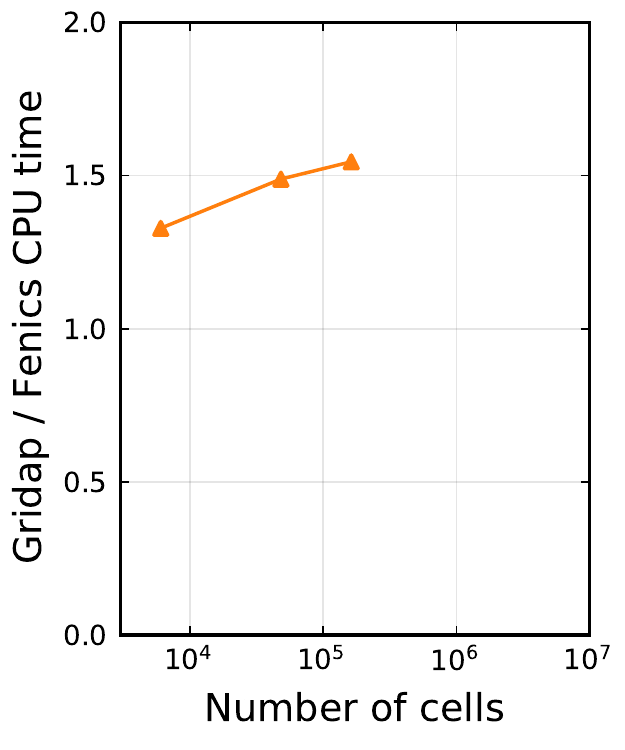}
\caption{In-place assembly with structured mesh.}
\end{subfigure}
\caption{Stokes benchmark: Scaling of CPU time for the phases \emph{assembly from scratch} and \emph{in-place assembly} for computations on unstructured mesh and structured meshes.}
\label{fig:stokes_cmp_times}
\end{figure}

This benchmark is run with Gridap and FEniCS. The Gridap driver is a variation of listing  \lst{lst:stokes_driver} in which we include timing routines and impose different boundary conditions. Again, the FEniCS code is based on the examples on the project web page. 
 As in the Poisson benchmark, we measure the time to assemble the linear system from scratch and to re-assemble the system reusing pre-computed information. The results are shown in \fig{fig:stokes_cmp_times}. Note that Gridap is significantly faster for assembly from scratch whereas FEniCS is slightly faster for in-place assembly. This time difference can be attributed to the more general implementation of the assembly loop for different cell topologies available in Gridap.

\section{Conclusions}
\label{sec:conclusions}

In this work, we have presented the main software abstractions behind the Gridap library. They enable the implementation of general-purpose \ac{fe} codes by leveraging the new possibilities provided by the emerging Julia programming language. We have detailed the main building blocks of the low-level computational back-end based on the multiple-dispatch paradigm of Julia, as well as the high-level  \ac{api} that allows the user to specify the weak form in a syntax almost identical to the whiteboard mathematical notation. 

The low-level back-end is extensible and highly customizable, whereas the high-level \ac{api} provides a convenient way to write \ac{pde} solvers in few lines of code. Both the back-end and front-end are implemented in the same programming language. We do not consider any compiler of variational forms or any code generation mechanism at the package level.  This feature is not usual in previous \ac{fe} frameworks with similar user \acp{api}. It enormously simplifies the usage, development and maintenance of the library. Numerical experiments for the Poisson and Stokes problems show that this new software design and the adoption of the Julia programming language lead to run-time performance similar to existing libraries like FEniCS. 

Due to length constraints, we have focused on the \ac{fe} discretization of linear single-field and multi-field problems, which is also the core functionality needed to solve nonlinear and transient problems. In any case, Gridap also provides a high-level interface for nonlinear problems, and time-dependent \acp{pde} can be conveniently solved with the GridapODEs extension package \cite{gridapodes_gh}. The functionality presented in this work can readily be used to integrate the local portions of distributed sparse systems of linear equations in parallel computations. A user \ac{api} for distributed-memory computations based on MPI will be soon released in the extension Package GridapDistributed \cite{gridapdistributed_gh}.

\section{Acknowledgments}

This research was partially funded by the Australian Government through the Australian Research Council (project number DP210103092), the European Commission under the FET-HPC ExaQUte project (Grant agreement ID: 800898) within the Horizon 2020 Framework Programme and the project RTI2018-096898-B-I00 from the ``FEDER/Ministerio de Ciencia e Innovación – Agencia Estatal de Investigación''. F. Verdugo acknowledges support from the Spanish Ministry of Economy and Competitiveness through the ``Severo Ochoa Programme for Centers of Excellence in R\&D (CEX2018-000797-S)".

\setlength{\bibsep}{0.0ex plus 0.00ex}
\bibliographystyle{unsrtnat}
\bibliography{refs}

\end{document}